# Crystal structure and magnetic modulation in $\beta$-Ce$_2$O$_2$FeSe$_2$

Chun-Hai Wang[1,2], C. M. Ainsworth[1], S.D. Champion[1], G.A. Stewart[3], M. C. Worsdale[4], T. Lancaster[4], S. J. Blundell[5], Helen E. A. Brand[6], and John S. O. Evans[1]

[1]*Department of Chemistry, Durham University, University Science Site, South Road, Durham, DH1 3LE, UK*
[2] *School of Chemistry, The University of Sydney, Sydney, NSW 2006, Australia*
[3]*School of Physical, Environmental & Mathematical Sciences, UNSW Canberra, Australian Defence Force Academy, PO Box 7916, Canberra, BC 2610, Australia*
[4]*Department of Physics, Durham University, University Science Site, South Road, Durham, DH1 3LE, UK*
[5]*Department of Physics, Oxford University, Clarendon Laboratory, Parks Road, Oxford, OX1 3PU, UK*
[6]*Australian Synchrotron, 800 Blackburn Rd., Clayton, Victoria, 3168, Australia*

## ABSTRACT

We report a combination of X-ray and neutron diffraction studies, Mössbauer spectroscopy and muon spin relaxation ($\mu^+$SR) measurements to probe the structure and magnetic properties of the semiconducting $\beta$-Ce$_2$O$_2$FeSe$_2$ oxychalcogenide. We report a new structural description in space group P$na2_1$ which is consistent with diffraction data and second harmonic generation measurements and reveal an order-disorder transition on one Fe site at $T_{OD} \approx 330$ K. Susceptibility measurements, Mössbauer and $\mu^+$SR reveal antiferromagnetic ordering below $T_N = 86$ K and more complex short range order above this temperature. 12 K neutron diffraction data reveal a modulated magnetic structure with **q** = 0.444 **b**$_N$*.

## I. INTRODUCTION

There has been significant recent research on oxychalcogenide materials due to their important electronic and magnetic properties and the area has been recently reviewed [1-2]. One important family of compounds is those with general composition Ln$_2$O$_2$MSe$_2$ (Ln = La, Ce; M = Mn, Fe, Zn, Cd) [3-11] and closely related composition such as La$_2$O$_2$Cu$_{2-4x}$Cd$_{2x}$Se$_2$ [12]. These materials are semiconductors with band gap varying from ~0.3 eV to ~ 3.3 eV [1, 3-5, 8, 10-11, 13-16]. For example, layered La$_2$O$_2$CdSe$_2$ was reported as a wide-gap (3.3 eV) semiconductor and investigated as an optoelectronic device component [14-15]. $\beta$-La$_2$O$_2$FeSe$_2$ and $\beta$-La$_2$O$_2$MnSe$_2$ are semiconductors with band gaps of 0.7 eV and 1.6 eV respectively [3]. These mixed-anion compounds have relatively flexible atomic interactions and can adopt different structure types (polymorphs). Three basic structure types have been observed for Ln$_2$O$_2$MSe$_2$ compositions to date. Most adopt a layered structure ($\alpha$-phase) which can be modulated in either a commensurate or incommensurate manner within individual layers to accommodate different transition metal arrangements [8, 10-11]. There are also two non-layered structures reported (orthorhombic $\beta$-phase and monoclinic Pb$_2$HgCl$_2$O$_2$-type $\gamma$-phase) [3, 13]. All three structures can be adopted by Ce$_2$O$_2$FeSe$_2$ by modifying the synthesis conditions [4, 13]. McCabe *et al*. reported the nuclear and magnetic structure of bulk layered $\alpha$-Ce$_2$O$_2$FeSe$_2$ [4, 6]; the space group of the nuclear structure is I$mcb$ (72). Nitsche *et al*. observed all three polymorphs (which they label as *oI-*, *oA-* and *mC-* Ce$_2$O$_2$FeSe$_2$ according to their symmetry) in single crystals grown at different temperatures [13]. The space groups of $\beta$- and $\gamma$- Ce$_2$O$_2$FeSe$_2$ they reported were A$mam$ (#63) and C2/$m$ (#12), respectively. We have investigated the bulk and single crystal forms of $\beta$-Ce$_2$O$_2$FeSe$_2$ and found that the diffraction patterns we observed disagree with the reflection conditions of space group A$mam$. We propose a different room temperature structural model for $\beta$-Ce$_2$O$_2$FeSe$_2$ based on our x-ray and neutron diffraction data and reveal an order-disorder transition involving one Fe site in the material at $T_{OD} \approx 330$ K. Low temperature neutron diffraction, Mössbauer spectroscopy and muon spin relaxation spectra all show that $\beta$-Ce$_2$O$_2$FeSe$_2$ orders antiferromagnetically below $T_N = 86$ K. We report the modulated magnetic structure and probe short range ordering above $T_N$.

## II. EXPERIMENTAL DETAILS

*Synthesis.* A polycrystalline sample of $\beta$-$Ce_2O_2FeSe_2$ was prepared by solid state reaction. $CeO_2$ (99.99%, Alfa Aesar, heated at 1000 °C before use), Se (99.999%, Alfa Aesar), and Fe (99.9%, Sigma-Aldrich) were weighed and ground in an agate mortar and pestle in a stoichiometric ratio. The well-mixed powders were placed in an alumina crucible and sealed in a silica tube with a second alumina crucible filled with Ti powder (99.5%, Alfa Aesar, 5% molar excess) acting as oxygen getter (forming $TiO_2$). The tubes were evacuated to < $10^{-2}$ mbar before sealing. The sealed tubes were heated slowly to 600 °C and held for 24 h, then to 1000 °C and held for 48 h. After cooling, the samples were ground, resealed in silica tubes, and reheated at 1000 °C for 48 h. An essentially single phase (>99%) product was obtained.

$\beta$-$Ce_2O_2FeSe_2$ single crystals were prepared from stoichiometric amounts of $CeO_2$, Fe and Se, in a KCl flux (99%, Alfa Aesar, heated to 150 °C before use). The molar amount of KCl was ~10 times that of $Ce_2O_2FeSe_2$. The well-ground mixture (~0.8 g total charge) was placed into an alumina crucible and sealed with a second alumina crucible filled with Ti (5% molar excess) powder. The tube was then heated to 600 °C at 60 °C/h, held for 24 h; heated to 950 °C at 60 °C/h, held for 96 h, ramped to 850 °C at 60 °C/h and held for 96 h, cooled to 600 °C at 2 °C/h and finally cooled to room temperature at 100 °C/h. The reacted mixture was washed with water to remove KCl and dried with acetone. Black blade or plate-like single crystals were obtained.

*Powder diffraction.* Laboratory powder X-ray diffraction data for $\beta$-$Ce_2O_2FeSe_2$ were collected at room temperature (RT) from 8 - 140° $2\theta$ in reflection mode using a Bruker D8 powder diffractometer on samples sprinkled on zero-background Si wafers. Cu $K\alpha$ radiation (tube condition: 50 kV, 40 mA), variable divergence slits and a Lynxeye Si strip position detector (PSD) were used. For Rietveld-quality data a scan step of 0.02° and scan time of ca. 38 h were used. Variable temperature PXRD data (~2 K intervals, 20 minutes scans) on $Ce_2O_2FeSe_2$ were recorded between 13 K and 300 K (on cooling and warming) with temperature controlled by an Oxford Cryosystems PheniX cryostat. Data were also collected using Mo $K\alpha$ radiation between 100 and 450 K on a sample loaded in a 0.7 mm capillary with temperature controlled by an Oxford Cryosystems Cryostream 700 compact. Synchrotron PXRD data were collected on the powder diffraction beamline at the Australian synchrotron. The sample was loaded in a 0.3 mm capillary, and data collected using a Mythen microstrip detector from 1 – 81° $2\theta$ with a wavelength of 0.6354462(7) Å. To cover the gaps between detector modules, 2 datasets were collected with the detector offset by 0.5° and then merged to a single dataset using PDViPeR. Time-of-flight (TOF) powder neutron diffraction (PND) data were collected on a 3.6 g sample held in an 8 mm diameter vanadium can on the General Materials (GEM) Diffractometer at the ISIS facility of the Rutherford Appleton Laboratory (UK). Data were collected by six detector banks over data ranges of (TOF and d-spacing): PND_bank1 1.1 – 27 ms (1.5 – 36 Å); PND_bank-2 1.4 – 20 ms (0.9 – 14 Å); PND_bank-3 1.3 – 22 ms (0.45 – 7.7 Å); PND_bank-4 1.4 – 20 ms (0.39 – 4.0 Å); PND_bank-5 1.5 – 18 ms (0.22 – 2.7 Å); PND_bank-6 1.6 – 16 ms (0.18 – 1.8 Å). PND data were acquired at room temperature for 3.5 h and at 12 K for 2.5 h.

TOPAS Academic (TA) [17] was used for the combined Rietveld refinement of the PXRD and PND data. ISODISTORT [18] (http://stokes.byu.edu/iso/isotropy.php) was used to derive the low temperature magnetic structure. We note that the (correct) magnetic form factor of $Ce^{3+}$ appears to be consistently mislabelled as "$Ce^{2+}$" in *International Tables for Crystallography* (Volume C) [19] and consequently in the scattering factor databases of TA and other diffraction software [11]. Selected figures were drawn using Vesta [20].

*Single crystal X-ray diffraction (SXRD).* SXRD data of $\beta$-$Ce_2O_2FeSe_2$ (~ 0.11 mm × 0.16 mm × 0.04 mm crystal) were collected using a Bruker D8 VENTURE single crystal diffractometer at room temperature. A Microfocus Mo radiation source (0.71073 Å, 50 kV, 1 mA) and PHOTON 100 CMOS detector were used. A total of 1020 frames were recorded for 7 s/frame. Data were processed using Bruker APEX2 software, a numerical absorption correction based on the crystal geometry was applied and data were analysed using JANA2006 [21].

*Magnetic properties.* Zero-field-cooled (ZFC) and field-cooled (FC) magnetic susceptibility ($\chi$) of ~ 0.1 g $\beta$-$Ce_2O_2FeSe_2$

samples was measured using a Quantum Design SQUID magnetometer in the temperature range of 2 – 300 K in a 1000 Oe magnetic field.

*Second-harmonic generation (SHG)*. Powder SHG measurements were performed on a modified Kurtz-nonlinear optical (NLO) system using a pulsed Nd:YAG laser with a wavelength of 1064 nm. A detailed description of the equipment and methodology has been published [22]. Unsieved powders were placed in separate capillary tubes and no index matching fluid was used in any of the experiments. The SHG, i.e. 532 nm light, was collected in reflection and detected using a photomultiplier tube. A 532 nm narrow-bandpass interference filter was attached to the tube in order to detect the SHG light only. The SHG measurements were carried out using a 'single-shot' at low power density to avoid sample decomposition.

*Mössbauer Spectroscopy*. $^{57}$Fe Mössbauer spectra were recorded at UNSW Canberra as a function of temperature using a liquid cryogen bath cryostat. The commercial $^{57}$Co:Rh source (≈ 15 mCi) was mounted externally and oscillated with sinusoidal motion. Finely ground specimen material (≈ 20 mg cm$^{-2}$) was mixed with CB$_4$ filler material and sandwiched between beryllium discs. A standard α-Fe foil was employed at room temperature for calibration of the drive velocity.

*$\mu^+$SR spectra of β-Ce$_2$O$_2$FeSe$_2$.* Zero-field muon-spin relaxation (ZF $\mu^+$SR) measurements were made on a polycrystalline sample of β-Ce$_2$O$_2$FeSe$_2$ using the GPS instrument at the Swiss Muon Source (SµS), Paul Scherrer Institut, Villigen, Switzerland. In a $\mu^+$SR experiment [23] spin-polarized positive muons are stopped in a target sample, where the muon usually occupies an interstitial position in the structure. The observed property in the experiment is the time evolution of the muon spin polarization, the behaviour of which depends on the local magnetic field at the muon site. Each muon decays, with an average lifetime of 2.2 µs, into two neutrinos and a positron, the latter particle being emitted preferentially along the instantaneous direction of the muon spin. Recording the time dependence of the positron emission directions therefore allows the determination of the spin-polarization of the ensemble of implanted muons. In our experiments positrons are detected by detectors placed forward (F) and backward (B) of the initial muon polarization direction. Histograms $N_F(t)$ and $N_B(t)$ record the number of positrons detected in the two detectors as a function of time following the muon implantation. The quantity of interest is the decay positron asymmetry function, defined as

$$A(t) = \frac{N_F(t) - \alpha_{exp} N_B(t)}{N_F(t) + \alpha_{exp} N_B(t)}, \tag{1}$$

where $\alpha_{exp}$ is an experimental calibration constant. $A(t)$ is proportional to the spin polarization of the muon ensemble $P(t)$.

### III. RESULTS AND DISCUSSION

#### A. Structure of *β*-Ce$_2$O$_2$FeSe$_2$

The single crystal XRD (SXRD) data of β-Ce$_2$O$_2$FeSe$_2$ can be indexed using an orthorhombic unit cell with $a$ = 17.201(2) Å, $b$ = 16.293(1) Å, $c$ = 3.9686(4) Å. The reconstructed (0 $k$ $l$) and ($h$ 0 $l$) sections are shown as Fig. S1 in the supplementary information (SI). Extra reflections (relative intensity ~0.4%) are observed compared to the *A*-centered A*mam* (#63) space group reported by Nitsche *et al*. [13], which indicates that β-Ce$_2$O$_2$FeSe$_2$ adopts a primitive space group. By analogy to the similar systems β-La$_2$O$_2$MnSe$_2$ and β-La$_2$O$_2$FeSe$_2$ [3, 13], the symmetry of β-Ce$_2$O$_2$FeSe$_2$ could be P*nma* (#62) or P*na*2$_1$ (#33; P*n*2$_1$*a* in same cell setting as P*nma*). Room temperature second-harmonic generation measurements were performed on a modified Kurtz-nonlinear optical (NLO) system using a previously published methodology [22]. β-Ce$_2$O$_2$FeSe$_2$ gave a comparable SHG signal to quartz (×1.2) indicating that the sample is non-centrosymmetric. The positive SHG result therefore suggests the non-centrosymmetric P*na*2$_1$ group.

Structure refinement in the two space groups gave no significant difference in fit between the models ($R_w$ = 6.94% and 6.90 % respectively). Note that the *b* and *c* axis in the two models are swapped to keep standard space group choices. Combined refinement of laboratory and sycnhrotron X-ray and neutron powder data were conducted with cell parameters, atomic coordinates, site occupancies of Fe2 and Fe3, and isotropic atomic displacement parameters (ADP) refined. There is again only a marginal improvement in fit using P$na2_1$, with the overall $R_{wp}$ changed from 2.03% in P$nma$ (30 fractional atomic coordinates refined) model to 2.02% in P$na2_1$ (44 fractional atomic coordinates). Structure parameters of the P$nma$ model are given in Table 1 and the P$na2_1$ model in SI; full details can be found from in SI CIF files.

The refined structure is shown in Figure 1 and contains building blocks which are familiar from other oxychalcogenide structures. Oxide ions are located in fluorite-like infinite ribbons built from four OCe$_4$ or OCe$_3$Fe1 edge-shared tetrahedra which run parallel to the *b* axis of the P$nma$ cell or the *c* axis of the P$na2_1$ cell (for comparison with the magnetic structure we will use the P$nma$ cell from here onwards). Each Ce site has 4 short bonds to the tetrahedrally coordinated O atoms and 4 longer bonds to Se in a distorted square anti-prismatic arrangement common to many mixed anion materials. The ribbon edges are terminated by the Fe1 site, the coordination environment of which is completed by 4 Se ions to give infinite chains of edge sharing Fe1O$_2$Se$_4$ octahedra which also run parallel to the P$nma$ *b* axis. Fe2 sits close to a site that would be trigonally prismatically coordinated by 5 Se atoms, however this site is better considered as two closely-separated face-sharing tetrahedral sites (Fe2 and Fe3). At *T* > 330 K the Fe is randomly disordered over these two sites whereas at room temperature there is partial or local ordering to a ~0.8:0.2 ratio of Fe2:Fe3. The Fe2 tetrahedra form infinite corner sharing chains parallel to *b*. There are some analogies to the structures of LnOFeAs-derived superconductors and LnOMSe$_2$ materials which have extended 2D layers of edge shared O$_4$Ln tetrahedra separated by FeAs$_4$ or FeSe$_4$ tetrahedral layers.[8] In *β*-Ce$_2$O$_2$FeSe$_2$ we can consider corrugated oxide-containing layers in the *ab* plane, though these layers contain both Ln and Fe; these layers are separated by corrugated Fe/Se layers.

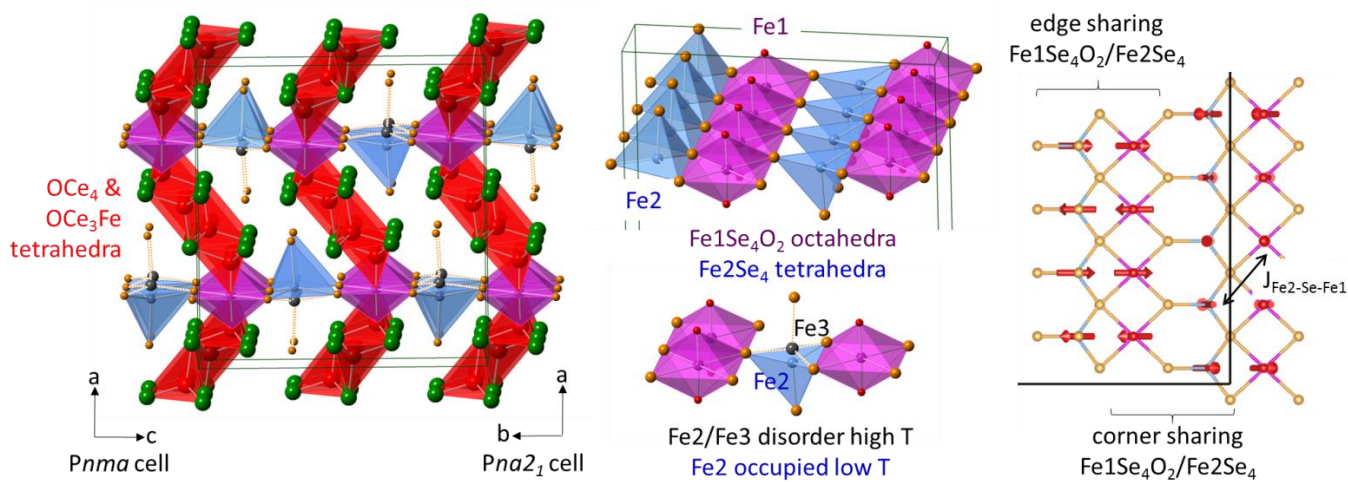

Figure 1. *β*-Ce$_2$O$_2$FeSe$_2$ structure viewed down either *b* axis of P$nma$ or *c* axis of P$na2_1$ cell. Middle views emphasise the chains of iron-centered polyhedra present and the relationship between the Fe2/Fe3 trigonal prismatic/paired tetrahedral sites. Right hand view shows Fe chains and moments (red arrows) from the best magnetic model.

Table 1. Structure parameters of β-Ce$_2$O$_2$FeSe$_2$ from combined refinement using PXRD and PND data (P*nma* model).

| | Space Group | P*nma* (62) | | | | |
|---|---|---|---|---|---|---|
| | $a$ (Å) | 17.18613(2) | | | | |
| | $b$ (Å) | 3.962980(5) | | | | |
| | $c$ (Å) | 16.28509(2) | | | | |
| | $V$ (Å) | 1109.150(3) | | | | |
| $d_{theory}$ (g/cm$^3$) | | 6.29987(2) | | | | |
| $R_{wp}$ (%) | | 2.03 (overall), 3.69 (lab X-ray), 3.06 (synchrotron X-ray), 3.68 (PND_bank1), 2.81 (PND_bank2), 1.83 (PND_bank3), 1.51 (PND_bank4), 1.43 (PND_bank5), 1.69 (PND_bank6). | | | | |
| | Site | $x$ | $y$ | $z$ | Occupancy | $B_{eq}$ (Å$^2$) |
| Ce1 | 4$c$ | 0.09063(8) | 0.25 | -0.01878(10) | 1 | 0.58(3) |
| Ce2 | 4$c$ | 0.40747(9) | 0.25 | -0.02259(10) | 1 | 0.56(3) |
| Ce3 | 4$c$ | 0.08679(7) | 0.25 | 0.67399(12) | 1 | 0.57(3) |
| Ce4 | 4$c$ | 0.41591(7) | 0.25 | 0.67681(12) | 1 | 0.73(3) |
| Se1 | 4$c$ | 0.24979(10) | 0.25 | 0.48740(3) | 1 | 0.869(16) |
| Se2 | 4$c$ | 0.06499(6) | 0.25 | 0.32808(9) | 1 | 0.84(2) |
| Se3 | 4$c$ | 0.42634(6) | 0.25 | 0.32959(10) | 1 | 0.92(2) |
| Se4 | 4$c$ | 0.25077(9) | 0.25 | 0.74886(3) | 1 | 0.922(16) |
| O1 | 4$c$ | 0.13800(9) | 0.25 | 0.12183(12) | 1 | 0.67(4) |
| O2 | 4$c$ | 0.35992(9) | 0.25 | 0.11954(11) | 1 | 0.49(4) |
| O3 | 4$c$ | 0.45455(11) | 0.25 | 0.54074(13) | 1 | 0.84(4) |
| O4 | 4$c$ | 0.04588(11) | 0.25 | 0.53961(13) | 1 | 0.64(4) |
| Fe1 | 4$c$ | 0.24960(9) | 0.25 | 0.11788(3) | 1 | 1.192(15) |
| Fe2 | 4$c$ | 0.27628(10) | 0.25 | 0.33753(9) | 0.788(3) | 1.004(18) |
| Fe3 | 4$c$ | 0.2366(3) | 0.25 | 0.3404(3) | 0.212(3) | 1.004(18) |

### B. Structural phase transition of β-Ce$_2$O$_2$FeSe$_2$

Variable temperature powder diffraction measurements were recorded for β-Ce$_2$O$_2$FeSe$_2$ to investigate possible phase transition beyond the magnetic ordering discussed below. Approximately 60 data sets recorded between 12 and 300 K were analysed and showed smooth and reversible behaviour on cooling and warming. Fractional cell parameter changes ($a$, $b$, $c$ and volume) are plotted in Figure 2, original values are given in SI. $b$, $c$ and volume $V$ increase as expected on warming but $a$ shows a local maximum followed by contraction at $T \approx 230$ K. This indicates a gradual phase transition, which, by analogy with related materials, is caused by ordering of Fe between Fe2 and Fe3 sites [3]. Lower quality diffraction data were recorded between 100 and 450 K and showed the phase transition is complete by ~330 K. The room temperature cell parameter is consistent with the high degree of Fe2 ordering refined from room temperature single crystal diffraction data, and its evolution on cooling suggests full ordering at low temperature. Further details will be discussed in the following sections. The low temperature cell parameters can be described by a single term Einstein-type expression [24],

$$\ln\left(\frac{x}{x_0}\right) = \frac{C\theta}{\exp(\theta/T) - 1} \qquad (2)$$

where $T$ is temperature, $x$ and $x_0$ are the cell parameters at $T$ and 0 K, $C$ a constant and $\theta$ the empirical "Einstein" temperature. The fitted 0 K cell parameters are $a_0 = 17.1738(9)$ Å (12 – 140 K), $b_0 = 3.9542(3)$ Å, $c_0 = 16.2290(7)$ Å, $V_0 = 1102.12(1)$ Å$^3$ and constants are $C_a = 1.0(1) \times 10^{-5}$ K$^{-1}$ (12 – 140 K), $C_b = 0.8(1) \times 10^{-5}$ K$^{-1}$, $C_c = 1.09(8) \times 10^{-5}$ K$^{-1}$, $C_V = 2.80(2) \times 10^{-5}$ K$^{-1}$ using a single "Einstein" temperature of 154(2) K. The fitted curves are shown in each figure.

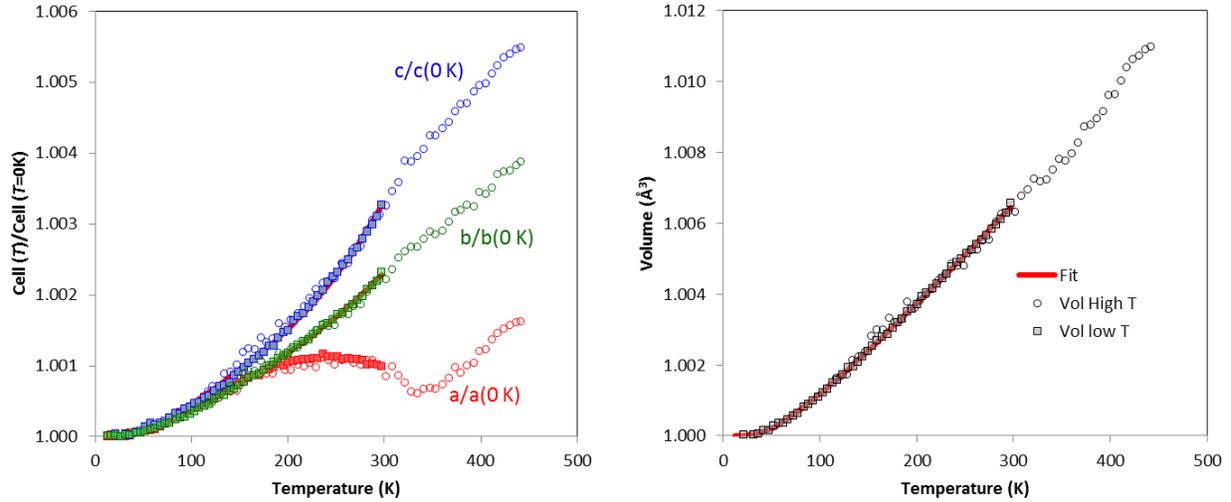

Figure 2. Cell parameter changes in $\beta$-Ce$_2$O$_2$FeSe$_2$ as a function of temperature. The solid red curves were fitted based on Eq. 2. Closed data points collected in Phenix cryostat in Bragg-Brentano mode; open data points using a capillary set up and Oxford cyrostream.

### C. 12 K Magnetic structure of $\beta$-Ce$_2$O$_2$FeSe$_2$

Extra peaks were observed in the 12 K PND data of $\beta$-Ce$_2$O$_2$FeSe$_2$ which arise from magnetic ordering and cannot be indexed using the nuclear unit cell. These peaks are most obvious in PND bank 3, which is plotted in Figure 3. The magnetic peaks can be indexed using the incommensurate magnetic ordering vector $\mathbf{q} = 0.444(1)\ \mathbf{b}_N^*$ based on an orthorhombic nuclear cell with $a_N = 17.607(3)$ Å, $b_N = 3.9624(7)$ Å, $c_N = 16.266(3)$ Å. The magnetic structure can therefore be described using a (3+1)D superspace model. However, as $\mathbf{q} \approx 4/9\ \mathbf{b}_N^*$ we can also use a supercell approximation using a cell nine times that of the nuclear cell in the $b$ direction. As discussed below, there are two magnetic modulation waves, which makes the supercell approach more straightforward.

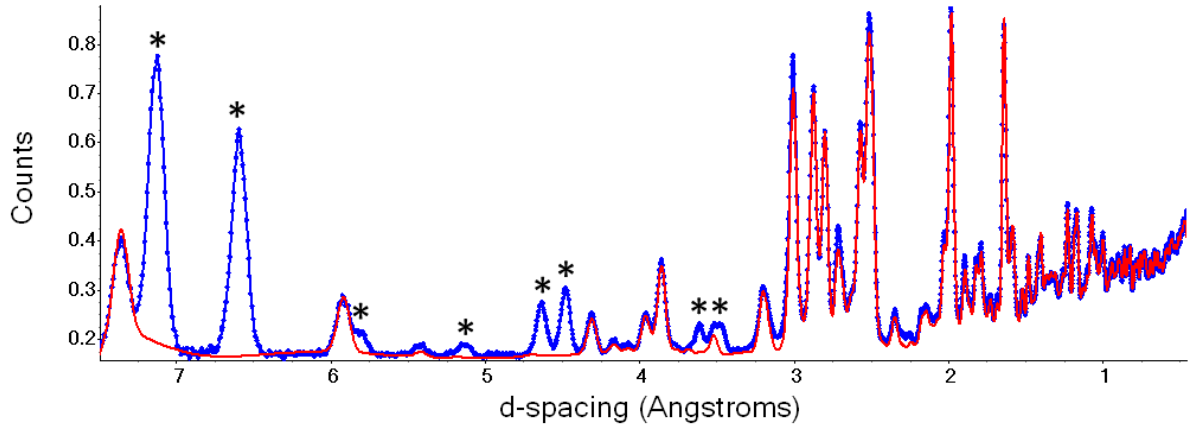

Figure 3. PND data (bank 3) of $\beta$-Ce$_2$O$_2$FeSe$_2$ at 12 K. Circles: experimental data; red solid line: simulated from nuclear model; peaks marked by stars come from the magnetic ordering.

To simplify the development of a model for the magnetic structure we made four initial assumptions: (1) As the structural difference between the P$nma$ and non-centrosymmetric P$na2_1$ (P$n2_1a$) models are minimal, we analyse and discuss the magnetic structure based on the centrosymmetric P$nma$ model with nuclear coordinates fixed at their room temperature values. (2) Since the magnetic diffraction is dominated by $Fe^{2+}$ moments, we only considered the $Fe^{2+}$ contributions during the development of different models. (3) Since the Fe2 site is close to fully-occupied at room temperature and orders further on cooling (Figure 2 and SI), we assumed full occupancy in our initial analysis; subsequent tests showed no improvement to fits on including minor Fe3 occupancy. (4) Ce contributions to the magnetic scattering were considered only for the best Fe-based models.

Based on these assumptions, magnetic structural models were derived using irrep analysis in the ISODISTORT suite. If there is one magnetic ordering vector mΔ (0 4/9 0) then there are four irrep possibilities: mΔ1, mΔ2, mΔ3, and mΔ4. This gives 12 possible magnetic structures depending on the phase shift of the magnetic modulation waves (origin shift). Figure 4 shows Rietveld fits of two of the best models with only $Fe^{2+}$ magnetic ordering (twelve parameters possible for each model, though not all are necessary to fit the data) considered: mΔ3 (33.147 P$na'2_1'$) and mΔ4 (33.146 P$n'a2_1'$); these gave similar $R_{wp}$ factors (2.52% vs 2.50% for all data compared to 6.05% with no magnetic contribution, Table 2 models 2 and 3). Both models give a reasonable description of the magnetic reflections. There are three types of modes for mΔ3 or mΔ4: $A'$, $A_1''$, and $A_2''$. Non-zero amplitudes of $A'$ modes give $Fe^{2+}$ moments parallel to the $b_N$ axis (or the Fe chains in Figure 4) whilst the $A_1''$ and $A_2''$ modes give moments parallel to $a_N$ or $c_N$ axis. In the mΔ3 model the dominant magnetic ordering is due to mΔ3-$A_2''$ modes with moments parallel to the $c_N$ axis, and in the mΔ4 model due to mΔ4-$A_1''$ modes with moments parallel to the $a_N$ axis. However, if we compare the relative intensities of peaks between 3.3 and 6.3 Å, we find that they are somewhat complementarity between mΔ3 and mΔ4 models (peaks over-calculated in one model are under-calculated in the other and *vice-versa*). This suggests that the magnetic structure might be a two-vector one (mΔ3 + mΔ4), containing contributions from both mΔ3 and mΔ4 magnetic ordering. Similar complementary is also observed in the other data banks.

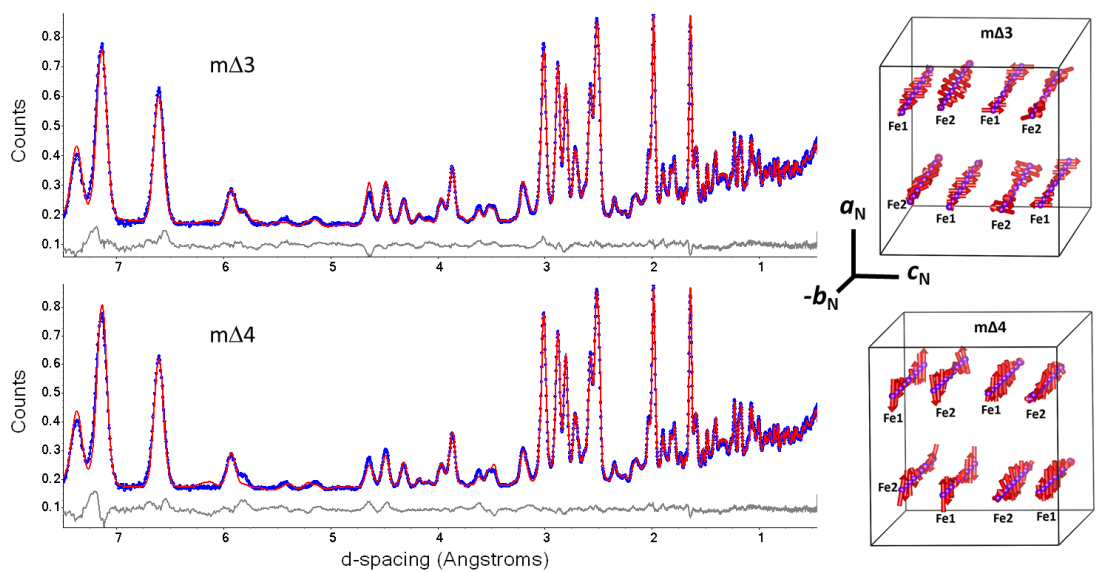

Figure 4. Rietveld fits to PND-bank 3 data of *β*-Ce$_2$O$_2$FeSe$_2$ at 12 K using mΔ3 and mΔ4 models (shown to the right of each curve) which correspond to space groups P$na'2_1'$ P$n'a2_1'$. Circles: experiment data; red solid line: calculated; only $Fe^{2+}$ moment considered. $R_{wp}$: 2.36% for mΔ3 and 2.45% for mΔ4.

Using two-vector mΔ3 + mΔ4 models, a better fit can be achieved using only two moment-defining parameters (mΔ3-$A_2''$ + mΔ4-$A_1''$ with Fe1, Fe2 having the same mΔ3-$A_2''$ or mΔ4-$A_1''$ amplitude) than using twelve in one-vector models (mΔ3 or mΔ4). With this model the $R_{wp}$ factor decreased to 2.29% for all data and 2.15% for PND-bank3 (Table 2, model 4). These refinements confirmed that the $Fe^{2+}$ moments are mainly in the $a_Nc_N$ plane. A further improvement in fit could be achieved by allowing $Ce^{3+}$ moments to refine (refined to be mainly along $b_N$ axis, see the cif files in SI and the discussion below), with $R_{wp}$ (overall) = 2.12% and $R_{wp}$ (PND-bank3) = 1.80% (Table 2, model 5). The observation of Ce ordering is consistent with several other Ce-Fe oxyselenides, where ordered Ce moments are required to fit the diffraction data and persist to surprisingly high temperatures [4, 6-7, 11]. An excellent fit to the PND data is achieved using this type of model, and the refined profiles for all banks are shown in Figure 5. The magnetic structural models can be found in the SI CIF files.

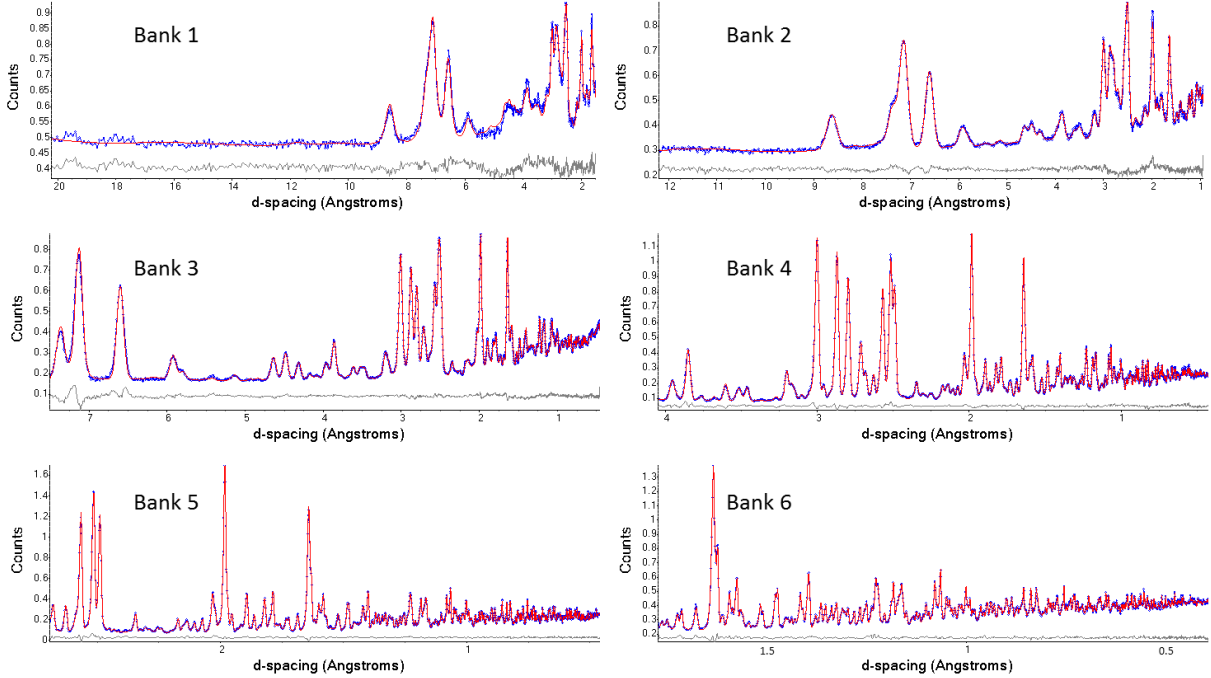

Figure 5. Rietveld-fitted PND data of $\beta$-$Ce_2O_2FeSe_2$ at 12 K using m$\Delta$3 + m$\Delta$4 models considering both $Fe^{2+}$ and $Ce^{3+}$ moments. Circles: experiment data; red solid line: calculated; Gray line: difference between calculated and experimental curve. $R_{wp}$ (bank 1): 2.41%; $R_{wp}$ (bank 2): 2.11%; $R_{wp}$ (bank 3): 1.80 %; $R_{wp}$ (bank 4): 2.36%; $R_{wp}$ (bank 5): 2.34%; $R_{wp}$ (bank 6): 1.51%.

If we consider the phase shift between m$\Delta$3 and m$\Delta$4 ordering, there are two coupling choices (phase choices with phase difference 0 or $\pi/2$). There are six combinations between $A_1''$ and $A_2''$ for m$\Delta$3 and m$\Delta$4 ordering for each choice. The way in which these determine the moments on the Fe1 chains is show in SI. The magnetic scattering is well described by the m$\Delta$3-$A_2''$ + m$\Delta$4-$A_1''$ model with either phase choice, which gave $R_{wp}$ values that differ by less than 0.01%. As such we cannot distinguish them from our refinements even though they are physically different. The refined magnetic moments of $Fe^{2+}$ for the two models are shown in Figure 6, where we introduce labels c1 to c4 to describe different chains derived from a single Fe1 or Fe2 parent site. In choice (a), the magnetic moments on the $Fe^{2+}$ chains form a planar amplitude modulated wave (spin density wave) with maximum moment ~5.19 $\mu_B$. However, in choice (b), the magnetic moment of $Fe^{2+}$ chains is mainly a direction modulated wave (helical proper screw) with approximately constant amplitude (moment from 3.30 to 4.04 $\mu_B$ along the chain). The refined moment from the choice (b) model is as expected for a high spin $Fe^{2+}$. The refined mode amplitudes of m$\Delta$3-$A_2''$ and m$\Delta$4-$A_1''$ [14(1):17(1)] are similar but not identical, which means a helical type ordering accompanied by a small modulation in spin amplitude. Along each $Fe^{2+}$ chain, the moment shows a local antiferromagnetic (AFM) like arrangement with the amplitude or direction modulated. Neighbouring Fe1 and Fe2 chains (Fe1c1-Fe2c1, Fe1c2-Fe2c2, Fe1c3-Fe2c3, and Fe1c4-Fe2c4) are aligned in a ferromagnetic (FM) sense whereas $Fe^{2+}$ chains c1-c4, c2-c3 are aligned AFM. For the modulation, $Fe^{2+}$ chains c1/c4 and c2/c3 have the same phase but there is anti-phase modulation between c1 (c4) and c2 (c3) chains.

Table 2. Summary of Rietveld Refinement Models.

| Model | Description | Magnetic Parameters | $R_{wp}$ (all banks) (%) | $R_{wp}$ (bank 3) (%) | gof |
|---|---|---|---|---|---|
| 1 | No magnetism | 0 | 6.05 | 8.05 | 6.54 |
| 2 | Fe m$\Delta$3 | 12×Fe | 2.52 | 2.37 | 2.71 |
| 3 | Fe m$\Delta$4 | 12×Fe | 2.50 | 2.45 | 2.69 |
| 4 | Fe (equated moments) m$\Delta$3+ m$\Delta$4 | 2×Fe | 2.29 | 2.15 | 2.47 |
| 5 | Fe+Ce m$\Delta$3+ m$\Delta$4 | 2×Fe + 8×Ce | 2.12 | 1.80 | 2.30 |

Although the $Ce^{3+}$ contribution to the magnetic scattering is weak, the data suggest that $Ce^{3+}$ magnetic moments are aligned parallel to the $Fe^{2+}$ chains ($b_N$ axis, m$\Delta$3-A' + m$\Delta$4-A') rather than along other axes (m$\Delta$3-A" + m$\Delta$4-A") [overall $R_{wp}$ values of 2.12% and 2.29% respectively]. Thus, the moment of $Ce^{3+}$ is described by the mode m$\Delta$3-A' and m$\Delta$4-A' (8 parameters to describe moments on the 4 $Ce^{3+}$ chains). The refined magnetic models (a) and (b) are shown in Figure 7. In both models, the moment of $Ce^{3+}$ shows the same phase shift as adjacent $Fe^{2+}$ which is consistent with $Ce^{3+}$ magnetic ordering being induced by $Fe^{2+}$. The relation between local $Ce^{3+}$ and $Fe^{2+}$ moments is counterintuitive and shows a "monopole" like behavior. A similar effect has been observed in $Ce_2O_2MnSe_2$ [11].

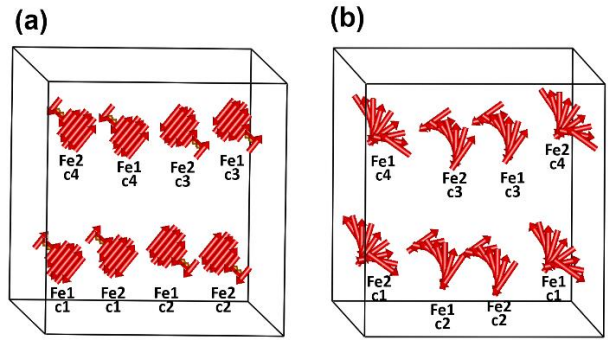

Figure 6. Alternate models for magnetic moment arrangement of $Fe^{2+}$. Fe1 & Fe2 chains refer to the Fe sites from *Pnma* nuclear structure and labels c1 – c4 define the chains which derive from the same reference nuclear Fe site.

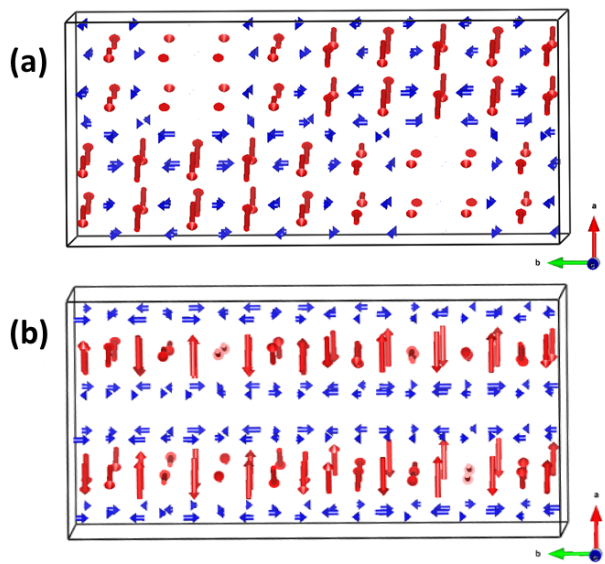

Figure 7. Alternate models for magnetic moment arrangement of $Fe^{2+}$ (red) and $Ce^{3+}$ (blue).

### D. Magnetic properties of *β*-Ce$_2$O$_2$FeSe$_2$

The temperature dependence of ZFC and FC molar magnetic susceptibilities ($\chi_{mol}$) of *β*-Ce$_2$O$_2$FeSe$_2$ are shown in Figure 8 along with the field dependence of moment at selected temperatures. For *β*-Ce$_2$O$_2$FeSe$_2$ the observed susceptibility can be reasonably approximated by the sum of contributions from Fe$^{2+}$ sites which order antiferromagnetically on cooling superimposed on a Curie-Weiss contribution from Ce$^{3+}$ which orders at a much lower temperature. This is consistent with our neutron and other observations. The Ce$^{3+}$ contribution makes estimation of $T_N$(Fe) from the magnetic data difficult, though we observe a sharp maximum around 80 K in $d\chi T/dT$ and a broader maximum around 145 K. The overall behavior is consistent with that of the (diamagnetic) La analogue *β*-La$_2$O$_2$FeSe$_2$, which shows non Curie-Weiss behavior with a broad hump in susceptibility around 91 K which coincides with the loss of magnetic neutron scattering at the Neel temperature $T_N$.

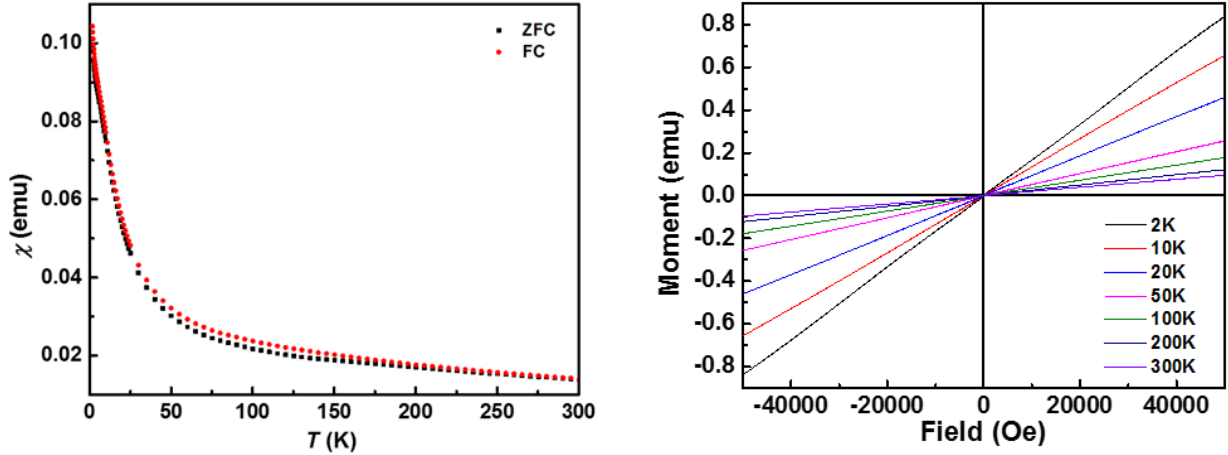

Figure 8. Magnetic properties of *β*-Ce$_2$O$_2$FeSe$_2$ as a function of temperature and field.

### E. Mőssbauer spectra of *β*-Ce$_2$O$_2$FeSe$_2$

$^{57}$Fe Mössbauer spectroscopy is a powerful probe of the magnetic properties of Fe-containing materials, and has been used to give important insight on various materials in which Fe orders incommensurately at low temperature. These include multiferroics, where cycloidal ordering often emerges from collinear sinusoidal ordering (e.g. FeVO$_4$ [25], BiFeO$_3$ [26-27] and AgFeO$_2$ [28]), iron arsenide superconductors (e.g. doped BaFe$_2$As$_2$ [29-30]) and systems such as Fe$_x$V$_{3-x}$S$_4$ [31], FeP [32], CuFeSe$_2$ and CuFeTe$_2$ [33]. Representative $^{57}$Fe Mössbauer spectra recorded for *β*-Ce$_2$O$_2$FeSe$_2$ between 5 and 300 K are shown in Figure 9. Spectra recorded above 90 K are typical of paramagnetic behavior and are a superposition of two symmetric, quadrupole-split doublets with line widths approaching instrument resolution. The intensities of the two doublets are equal below 200 K, consistent with their assignment to the Fe1 and Fe2 4a-sites. At 90 K, there is evidence of magnetic line broadening, evolving to a complex superposition of magnetically-split sextets at 5 K. Coupled with the sharp maximum observed for $d\chi T/dT$ near 80 K and $\mu^+$SR data discussed below, this confirms that the Fe sub-lattices are magnetically ordered below $T_N$(Fe) ≈ 85–90 K.

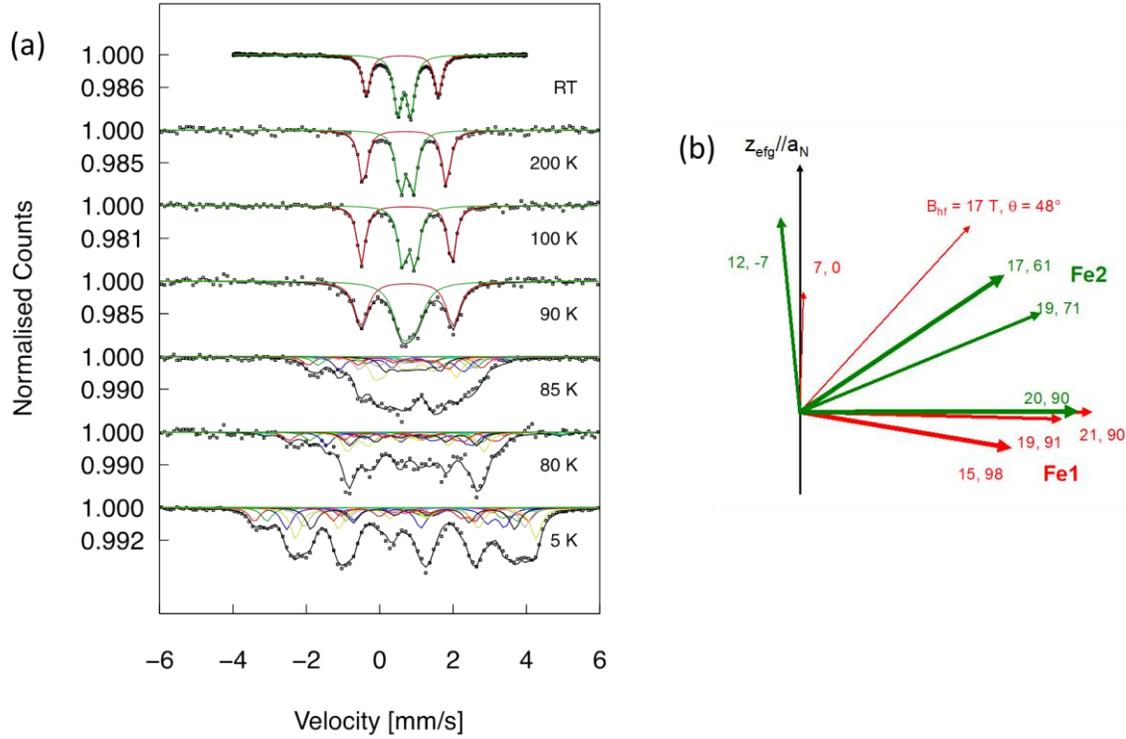

Figure 9. (a) Representative $^{57}$Fe-Mössbauer spectra recorded for β-Ce$_2$O$_2$FeSe$_2$ between 5 and 300 K. The solid lines show the fitted sub-spectra (colours) and their sum (black). (b) Diagrammatical representation of the magnetic hyperfine field vectors, $\boldsymbol{B}_{hf}$, fitted to the $T = 5$ K spectrum. The magnitude, orientation and thickness of the vectors represent the magnitude, polar angle $\theta$, and sub-spectrum intensity, respectively.

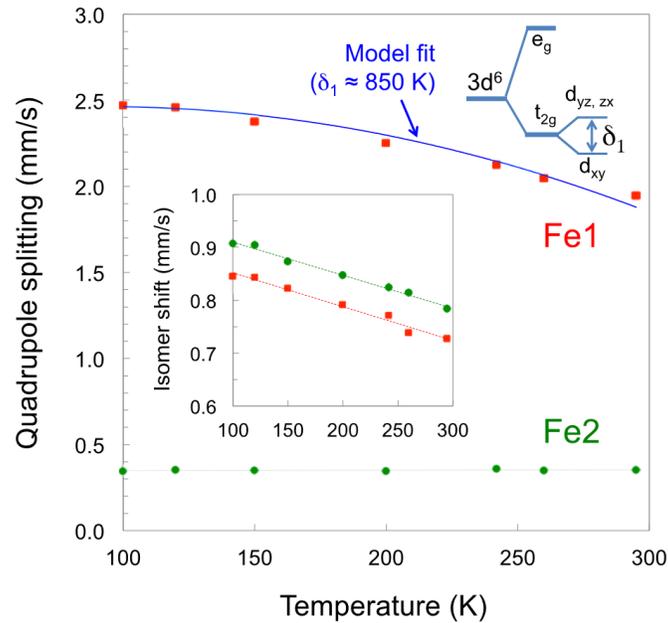

Figure 10. Experimental quadrupole splitting, $\Delta E_Q$ (main panel) and isomer shift, $\delta$ (lower inset) for high spin Fe$^{2+}$ at the Fe1 (solid red squares) and Fe2 (solid green circles) sites in β-Ce$_2$O$_2$FeSe$_2$. The blue theory curve is modelled on splitting of the $t_{2g}$ ground state (shown at top right) due to compressive, tetragonal distortion of the octahedral Fe1 site.

Table 3. Hyperfine interaction parameters fitted to $^{57}$Fe-Mössbauer spectra recorded for $\beta$-Ce$_2$O$_2$FeSe$_2$ at room temperature ($\delta$ = isomer shift relative to $\alpha$-Fe at room temperature, $\Delta E_Q$ = quadrupole splitting).

| Intensity (%) | $\delta$(re $\alpha$-Fe) (mm/s) | $\Delta E_Q$ (mm/s) | Site |
|---|---|---|---|
| 41.82(2) | 0.73(2) | 1.95(3) | Fe1 |
| 58.18(2) | 0.78(1) | 0.35(2) | Fe2 |

Values of the isomer shift, $\delta$, and quadrupole splitting, $\Delta E_Q$, derived by fitting to the paramagnetic spectra ($T$ > 90 K) are presented in Figure 10 as a function of temperature. Their room temperature values are also included in Table 3. The isomer shifts for the two doublets are similar and typical for high spin Fe$^{2+}$ ions in either octahedral or tetrahedral environments [34], but site assignment is made possible via their distinct quadrupole splitting behaviour.

The doublet with the larger $\Delta E_Q$ values (red fitted sub-spectra in Figure 9) can be assigned to the Fe1 octahedral site. Based on the structural data of Table 1 there is a mean bond length of 2.91(2) Å for the four selenium ligands in the $b_{NC_N}$ plane and 1.91(1) Å for the two apical oxygen ligands. This corresponds to a tetragonally-distorted octahedron that is compressed along the 4-fold axis through the O ligands. Under these circumstances, the low-lying octahedral $t_{2g}$ level is split into a singlet ($d_{xy}$) ground state and doublet ($d_{yz}$, $d_{zx}$) excited state. For high spin Fe$^{2+}$ and temperatures at which the spin-orbit coupling can be ignored, the thermal distribution of the sixth electron over these three states determines the temperature dependence of the principal electric-field gradient (efg) component acting at the $^{57}$Fe nucleus. The total quadrupole splitting is then expressed as [35-36]

$$\Delta E_Q(T) = \Delta E_Q(\text{latt}) + \Delta E_0 \cdot \left(1 - e^{-\delta_l k_B T}\right) / \left(1 + 2e^{-\delta_l k_B T}\right) \text{ with } \Delta E_0 = \frac{1}{2}e^2 Q \cdot \frac{\frac{4}{7}(1-R)\langle r^{-3}\rangle}{4\pi\varepsilon_0}, \quad (2)$$

where $\delta_l$ is the splitting of the $t_{2g}$ level, and $\Delta E_0 \approx +3.8$ mm/s (after de Grave *et al*. [37]) is the valence contribution to the quadrupole splitting at $T \to 0$ K (i.e. due to the $d_{xy}$ singlet), and $\Delta E_Q(\text{latt})$ is the contribution due to the charges on the surrounding lattice. Point charge model summations were employed to estimate $\Delta E_Q(\text{latt}) \approx -1.8$ mm/s, which is of *opposite sign* to the valence contribution. The experimental data were reasonably well described (fitted blue line in Figure 10) using $\Delta E_Q(\text{latt}) = -1.4$ mm/s, $\Delta E_0 = +3.8$ mm/s and $\delta_l \approx 850$ K.

The second doublet (green fitted sub-spectra in Figure 9) is then assigned to the Fe2 tetrahedral site. In this case, $\Delta E_Q$ is substantially smaller and essentially temperature independent. The room temperature combination of a high isomer shift ($\delta \approx$ 0.8 mm/s) and low quadrupole splitting ($\Delta E_Q \approx 0.35$ mm/s) is seemingly rare in the literature. However, similar results have been reported for Fe$^{2+}$ located at the tetrahedral sites of binary oxides and chalcogenides. Examples include $\delta = 0.84(3)$ mm/s, $\Delta E_Q = 0.35(3)$ mm/s for impurity Fe$^{2+}$ implanted in single crystal, hexagonal ZnO [38] and ranges of $\delta \approx 0.4$–0.6 mm/s, $\Delta E_Q = 0.2$–0.3 mm/s with relatively small temperature dependence for FeSe [39-41], FeTe [40-41] and Fe$_{1-x}$Mn$_x$Se$_{0.85}$ [42]. In the case of the chalcogenides, the slightly smaller isomer shift value has led some authors to conclude that the Fe$^{2+}$ is in its low spin ($S = 0$) state. However, this is unlikely for the tetrahedral Fe2 site of $\beta$-Ce$_2$O$_2$FeSe$_2$, given that the neutron diffraction analysis assigns it to either a sinusoidal or spiral magnetic structure with a moment amplitude close to $\mu = g_J S = 2 \times 2 = 4 \mu_B$. The simple trigonally-distorted tetrahedral site model outlined by Gerard *et al*. [43-44] is relevant to our experimental observations. In that model, the trigonal distortion modifies the expansion coefficients of the degenerate, tetrahedral $e$ ground state doublet and splits the excited $t_2$ triplet into a singlet and a doublet. The modified ground state doublet coefficients lead to a temperature-independent $V_{zz}$ contribution that depends on $\delta/\Delta$ where $\Delta$ is the overall tetrahedral splitting energy and $\delta$ is the trigonal distortion splitting of the upper state. Typically, $\delta << \Delta$, so that this model offers qualitative support for the small, temperature-independent, quadrupole splitting observed here for the Fe2 site.

Below the $T_N \approx 85$ K magnetic transition, the $^{57}$Fe-Mössbauer spectra were initially fitted as a superposition of four magnetically-split sextets for each of the Fe1 and Fe2 sites. This approach was prompted by the incommensurate magnetic vector $\mathbf{q} = 0.444(1)\ \mathbf{b}_N^* \approx 4/9\ \mathbf{b}_N^*$ for which there are expected to be four or five distinct magnitudes [magnetic structure model (a)] or orientations [model (b)] of the magnetic hyperfine field, $\mathbf{B}_{hf}$. For each site, the isomer shift, $\delta$, and the quadrupole splitting value, $\frac{1}{2}eQV_{zz}$, were fixed at values extrapolated from the high-temperature, paramagnetic spectra. The principal $z$-axis of the electric-field gradient was assumed to align with the $a_N$-axis and the asymmetry parameter, $\eta$, was fixed at zero for the Fe1 site but allowed to vary for the less symmetric Fe2 site. Only the magnitudes and orientations (the polar angle, $\theta$, with respect to the principal $z$-axis) of the individual $\mathbf{B}_{hf}$ were allowed to vary. Within each sextet, the Lorentzian line widths were set at 0.3 mm/s and the relative line intensities were fixed at 3:2:1:1:2:3 as appropriate for random orientation of the specimen crystallites. The results for the 5 K spectrum are presented diagrammatically in Figure 9(b) where the magnitude, orientation and thickness of the vectors represent the magnitude, polar angle $\theta$, and intensity, respectively, of the sub-spectral $\mathbf{B}_{hf}$. There is evidently a broad grouping of the $\mathbf{B}_{hf}$ into sets centred on $\theta \approx 0°$, 60° and 95° with $\mathbf{B}_{hf}$ ranging from 12 to 20 T (corresponding to Fe moments ranging from 2.4 to 4 $\mu_B$). This points to the model (b) magnetic structure as the preferred incommensurate magnetic model. As discussed previously in the literature (for example by Colson et al. [25]), for an equal-moment helical or conical structure the magnetic sextets should all be fitted with the same (or at least very similar) $\mathbf{B}_{hf}$ values. Only the quadrupole interaction's contribution would be expected to vary, in this case via the polar angle $\theta$. The outcome would be a single magnetic sextet for each of the Fe1 and Fe2 sites, but with a characteristic line-dependent broadening. However, in the case of an additional variation in the local moment (such as the range of 3.3 - 4.0 $\mu_B$ found in model (b)) the spectra are expected to be more complex, approaching those for the elliptical helical structure described by Colson et al. (Figure 3 of [25]).

### F. $\mu^+$SR spectra of β-Ce$_2$O$_2$FeSe$_2$

Representative zero field $\mu^+$SR spectra measured on $\beta$-Ce$_2$O$_2$FeSe$_2$ are shown in Figure 11. In spectra measured below $T = 85$ K oscillations in the asymmetry are observed. These oscillations are characteristic of a quasi-static local magnetic field at the muon stopping site, which causes a coherent precession of the spins of those muons with a component of their spin polarization perpendicular to this local field (expected to be 2/3 of the total polarization for a powder). The frequency of the oscillations is given by $v_i = \gamma_\mu B_i/2\pi$, where $\gamma_\mu$ is the muon gyromagnetic ratio ($= 2\pi \times 135.5$ MHz T$^{-1}$) and $B_i$ is the average magnitude of the local magnetic field at the $i$th muon site. Any fluctuation in magnitude of these local fields will result in a relaxation of the oscillating signal, described by a relaxation rate $\lambda$. The presence of oscillations at low temperatures provides unambiguous evidence that $\beta$-Ce$_2$O$_2$FeSe$_2$ is magnetically ordered throughout its bulk below 85 K.

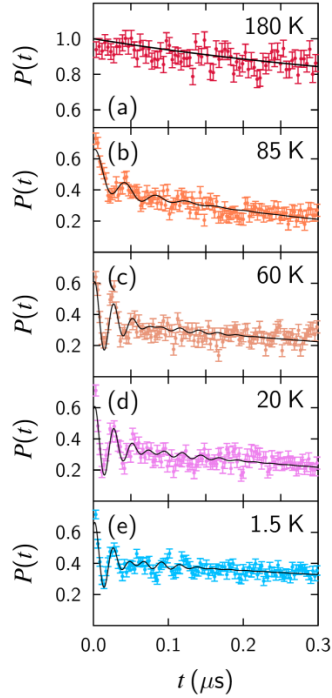

Figure 11. ZF $\mu^+$SR spectra measured at several temperatures.

The polarization spectra in the low temperature regime $T < 85$ K were found to be best fitted by the sum of two oscillating components with frequencies $\nu_1$ and $\nu_2$ and respective relaxation rates $\lambda_1$ and $\lambda_2$. The observation of two frequencies implies that muons stop at two magnetically distinct sites in the crystal. To model the data, we also require a third, purely exponential component with amplitude $P_{\text{bg}}$ and small relaxation rate $\lambda_3$ which accounts for those muons that are stopped in the sample holder or cryostat tails. We fit the data to the resulting function

$$P(t) = P_1 e^{-\lambda_1 t} \cos(2\pi \nu_1 t + \phi_1) + P_2 e^{-\lambda_2 t} \cos(2\pi \nu_2 t + \phi_2) + P_{\text{bg}} e^{-\lambda_3 t} \tag{5}$$

where $P(t) = A(t)/A_{\max}$ (Eq. 1), with $A_{\max}$ the maximum value of $A(t=0)$ observed in the paramagnetic phase (see below). The phase offsets were found to be constant at $\phi_1 = -15.8°$ and $\phi_2 = -19.5°$. The amplitudes of the oscillatory components were found to be $P_1 = 0.13$ and $P_2 = 0.15$, implying that the two magnetically distinct muon sites are occupied with similar probability.

The results of fitting Eq. 5 to the measured data are shown in Figure 12(a,b). We note that the two frequencies do not show an identical temperature dependence, and attempts to fit them in fixed proportion were unsuccessful. This might reflect a subtle change in the magnetic structure with temperature or, perhaps more likely, the difficulty in fitting the data consistently across a large temperature range in a case where the observed frequencies are large. The evolution of the relaxation rates $\lambda_{1,2}$ [Figure 12(b)] is also of possible significance and shows a local maximum at 10 K and the suggestion of a minimum around 40 K. The temperature evolution of the larger of the two frequencies [Figure 12(a)] was fitted, close to the transition, to the phenomenological form

$$\nu(t) = \nu(0)\left[1 - \frac{T}{T_{\text{N}}}\right]^{\beta}. \tag{6}$$

Several parametrizations are possible, but from the fits we estimate $T_{\text{N}} = 86 \pm 1$ K and $\beta \approx 0.25$.

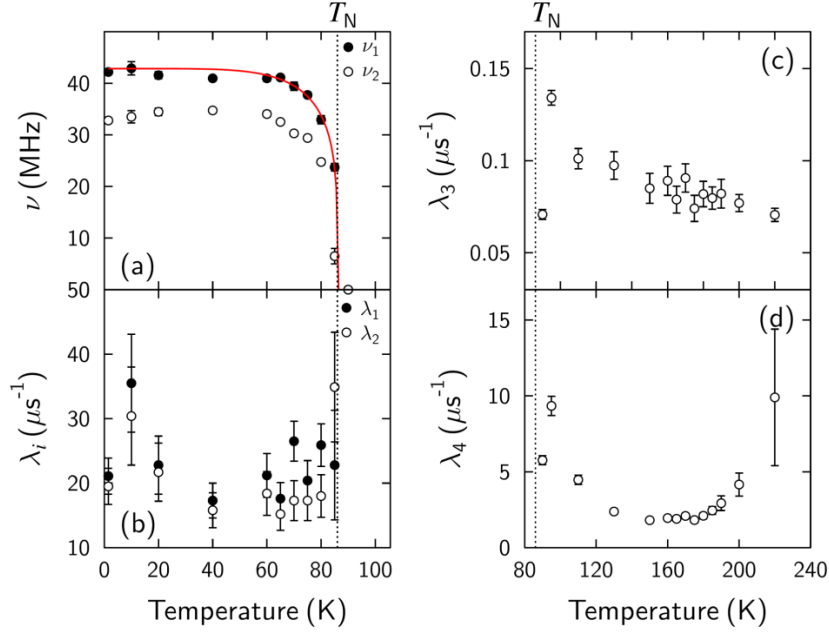

Figure 12. (a-b) Results of fitting the $\mu^+$SR spectra to Eq. 5 and (c,d) to Eq. 7. The line in (a) is a guide to the eye from Eq. 6.

Above $T_N$, the spectra show a monotonic decrease and demonstrate the system is in a magnetically disordered state with dynamic field fluctuations on the muon time scale. The data are most successfully fitted using the sum of two relaxing components with the function

$$P(t) = P_3 e^{-\lambda_3 t} + P_4 e^{-\lambda_4 t} + P_{\text{bg}}. \tag{7}$$

The results of fitting with relaxation rate $\lambda_4 \gg \lambda_3$, $P_3 = 0.66$ and $P_4 = 0.32$ are shown in Figure 12(c,d). The small relaxation rate $\lambda_3$ shows a steady decrease with $T$ but the larger relaxation rate $\lambda_4$ shows a distinct minimum around 150 K [Figure 12(d)], close to the small feature seen in the magnetic susceptibility. This is possibly suggestive of the muon seeing a crossover between two different regimes of magnetic behaviour, with distinct sets of relaxation process on either side (presumably corresponding to differences in dynamics, or the field distribution itself).

## IV. DISCUSSION/CONCLUSIONS

In conclusion we have used a combination of X-ray and neutron diffraction techniques to investigate the nuclear and magnetic structures of $\beta$-Ce$_2$O$_2$FeSe$_2$ and used Mössbauer and $\mu^+$SR techniques to probe the temperature dependence of its magnetic order. There is clear evidence from single crystal X-ray diffraction that the structure is primitive rather than centred at room temperature. The systematic absences and the observation of second harmonic generation suggest space group P$na2_1$. An excellent fit to both single crystal X-ray diffraction and powder X-ray and neutron diffraction data can be achieved with this space group. Variable temperature powder diffraction experiments show an order-disorder transition associated with the Fe2/Fe3 sites occurs at around 330 K. Above this temperature Fe is statistically disordered over two closely-separated face-sharing tetrahedral sites (thus appearing in a pseudo-trigonal bipyramidal site). At room temperature 80% site ordering is achieved, and ordering is essentially complete below 230 K.

The structure of $\beta$-Ce$_2$O$_2$FeSe$_2$ contains chains of edge-shared Fe1O$_2$Se$_4$ distorted octahedra and corner shared Fe2Se$_4$ tetrahedra with the tetrahedral and octahedral chains linked by either edge or corner sharing. Below $T_N = 86$ K Fe sites order antiferromagnetically within each chain to give an incommensurately modulated magnetic structure with $\mathbf{q} = 0.444\ \mathbf{b}_N^*$, which can be approximate using a 9-fold superstructure along the b-axis. From a visual comparison of diffraction data, $\beta$-$La_2O_2FeSe_2$ [3] appears to have a similar magnetic structure. It is difficult to be definitive about the exchange interactions leading to this complex magnetic structure from the data available. Similar $FeSe_4O_2$ edge sharing chains are observed in the $Ln_2O_2Fe_2OSe_2$ family of materials, though as part of infinite 2D layers made up of face sharing octahedra. In these materials crystalline electric field anisotropy leads to a preference for Fe moments along Fe–O bonds [45] and moments are ordered ferromagnetically along each edge-shared chains [consistent with Goodenough-Kanamori-Anderson (GKA) predictions]. The exchange constants are, however, relatively weak and both Mn and Co analogues are found computationally and experimentally to violate GKA rules and have antiferromagnetic order along the chains [46-48]. In $\beta$-$Ce_2O_2FeSe_2$ the local magnetic structure is probably governed by strong Fe1-Se-Fe2 ~165° antiferromagnetic exchange within the edge shared $Fe1Se_4O_2$/$Fe2Se_4$ double chains (see Figure 1), consistent with GKA predictions. Interchain coupling is more complex and presumably gives rise to the frustration leading to the incommensurate structure, though we note that incommensurate order can be observed even in the geometrically simpler $Ln_2F_2Fe_2OSe_2$ systems [49]. Mössbauer and $\mu^+$SR techniques have confirmed the low temperature magnetic order and suggest that the material has a modulated structure based on an elliptical proper screw. Both techniques suggest short range magnetic order is retained significantly above $T_N$.

## ACKNOWLEDGMENTS

We thank EPSRC for funding under EP/J011533/1 and EP/N024028/1. Powder diffraction data were collected on the Powder Diffraction beamline at the Australian synchrotron. We thank Ivana Evans, Matthew Tate, Nicola Scarlett and Garry McIntyre for assistance with data collections. JSOE would like to thank ANSTO for a visiting position during which part of this research was performed. We thank Professor Shiv Halasyamani of the University of Houston for SHG measurements. We thank F. Kirschner, H. Luetkens and A. Amato for experimental assistance with $\mu^+$SR experiments.

# Crystal structure and magnetic modulation in $\beta$-Ce$_2$O$_2$FeSe$_2$

Chun-Hai Wang[1,2], C. M. Ainsworth[1], S.D. Champion[1], G.A. Stewart[3], M. C. Worsdale[4], T. Lancaster[4], S. J. Blundell[5], Helen E. A. Brand[6], and John S. O. Evans[1]

[1]*Department of Chemistry, Durham University, University Science Site, South Road, Durham, DH1 3LE, UK*
[2]*School of Chemistry, The University of Sydney, Sydney, NSW 2006, Australia*
[3]*School of Physical, Environmental & Mathematical Sciences, UNSW Canberra, Australian Defence Force Academy, PO Box 7916, Canberra, BC 2610, Australia*
[4]*Department of Physics, Durham University, University Science Site, South Road, Durham, DH1 3LE, UK*
[5]*Department of Physics, Oxford University, Clarendon Laboratory, Parks Road, Oxford, OX1 3PU, UK*
[6]*Australian Synchrotron, 800 Blackburn Rd., Clayton, Victoria, 3168, Australia*

Table S1. Structure parameters of $\beta$-Ce$_2$O$_2$FeSe$_2$ from combined refinement using PXRD and PND data (P$na$2$_1$ model) [a].

| | | | | | | |
|---|---|---|---|---|---|---|
| Space Group | | P$na$2$_1$ (33) | | | | |
| $a$ (Å) | | 17.18613(2) | | | | |
| $b$ (Å) | | 16.28510(2) | | | | |
| $c$ (Å) | | 3.962979(5) | | | | |
| $V$ (Å) | | 1109.150(3) | | | | |
| $d_{\text{theory}}$ (g/cm$^3$) | | 6.29987(2) | | | | |
| $R_{\text{wp}}$ (%) | | 2.02 (overall), 3.69 (lab X-ray), 3.06 (synchrotron X-ray), 3.68 (PND_bank1), 2.81 (PND_bank2), 1.82 (PND_bank3), 1.51 (PND_bank4), 1.41 (PND_bank5), 1.65 (PND_bank6). | | | | |

| | Site | $x$ | $y$ | $z$ | Occupancy | |
|---|---|---|---|---|---|---|
| Ce1 | 4$a$ | 0.40755(9) | 0.02273(10) | 0.25 | 1 | 0.47(3) |
| Ce2 | 4$a$ | 0.41325(8) | 0.82595(12) | 0.752(6) | 1 | 0.47(3) |
| Ce3 | 4$a$ | 0.08419(7) | 0.82329(12) | 0.752(6) | 1 | 0.55(3) |
| Ce4 | 4$a$ | 0.09070(8) | 0.01866(10) | 0.248(8) | 1 | 0.39(3) |
| Se1 | 4$a$ | 0.24988(10) | 0.51258(3) | 0.249(4) | 1 | 0.749(16) |
| Se2 | 4$a$ | 0.42625(6) | 0.67052(10) | 0.251(5) | 1 | 0.77(2) |
| Se3 | 4$a$ | 0.06484(6) | 0.67184(9) | 0.253(6) | 1 | 0.72(2) |
| Se4 | 4$a$ | 0.24917(9) | 0.75113(3) | 0.754(4) | 1 | 0.813(16) |
| O1 | 4$a$ | 0.13804(9) | 0.87821(12) | 0.245(8) | 1 | 0.60(5) |
| O2 | 4$a$ | 0.35996(10) | 0.88037(11) | 0.247(8) | 1 | 0.31(4) |
| O3 | 4$a$ | 0.45405(11) | 0.96077(13) | 0.750(10) | 1 | 0.60(4) |
| O4 | 4$a$ | 0.04539(11) | 0.95885(13) | 0.749(10) | 1 | 0.67(4) |
| Fe1 | 4$a$ | 0.24967(9) | 0.88209(3) | 0.262(4) | 1 | 1.010(17) |
| Fe2 | 4$a$ | 0.27581(10) | 0.66217(10) | 0.251(7) | 0.797(3) | 0.92(2) |
| Fe3 | 4$a$ | 0.2352(3) | 0.6605(3) | 0.249(17) | 0.203(3) | 0.92(2) |

[a]: The $z$ coordinate of Ce1 was fixed to 0.25 to keep the center of mass near to that of P$nma$ model.

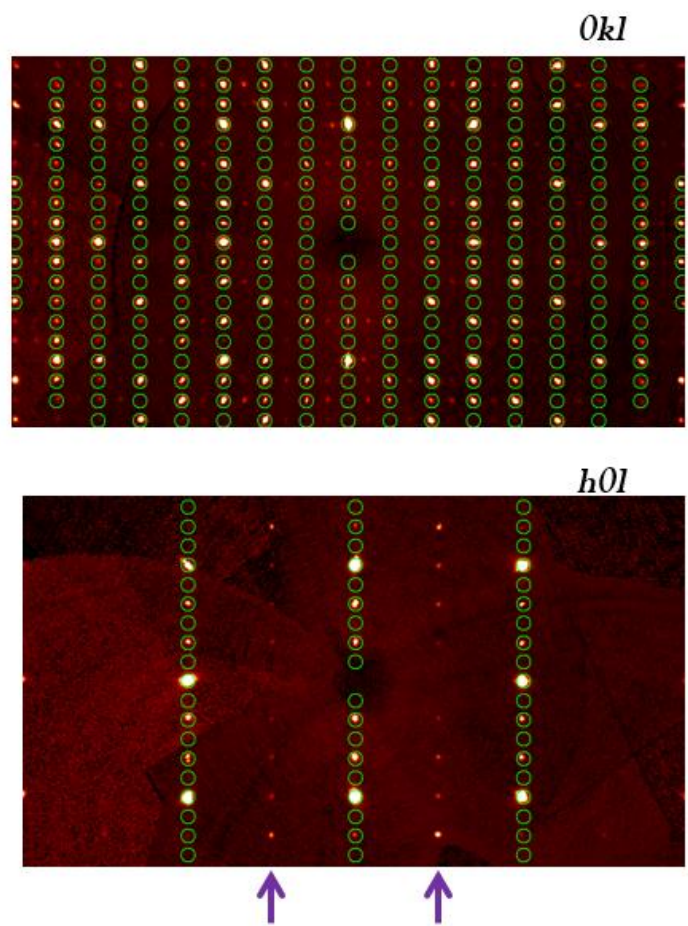

Fig. S1. Selected cross sections of single crystal XRD image of $\beta$-$Ce_2O_2FeSe_2$. Green circles are reflections predicted by *Amam* space group. Extra reflections pointed by arrow were observed.

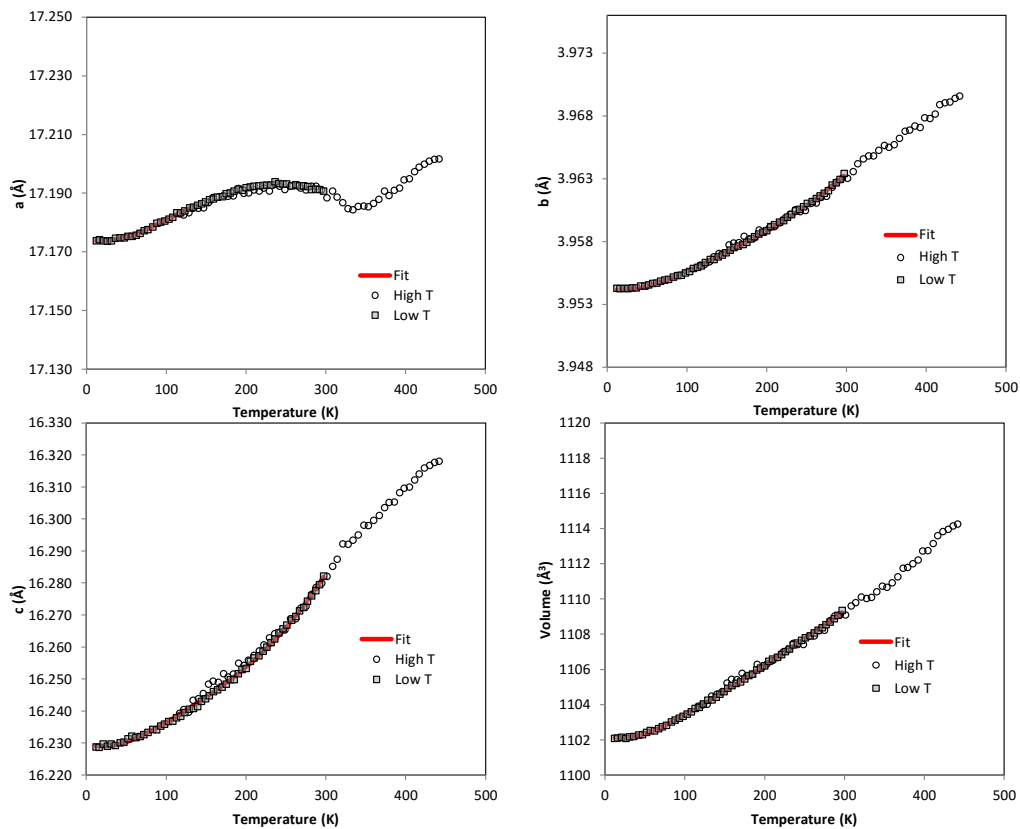

Figure S2. Temperature dependence of cell parameters of $\beta$-Ce$_2$O$_2$FeSe$_2$. The solid curves were fitted based on Eq. 2. Closed data points collected in Phenix cryostat in Bragg-Brentano mode; open data points using a capillary set up and Oxford cyrostream. Lower-quality capillary data have been offset by 0.009, 0.002 and 0.007Å for $a$, $b$ and $c$ respectively.

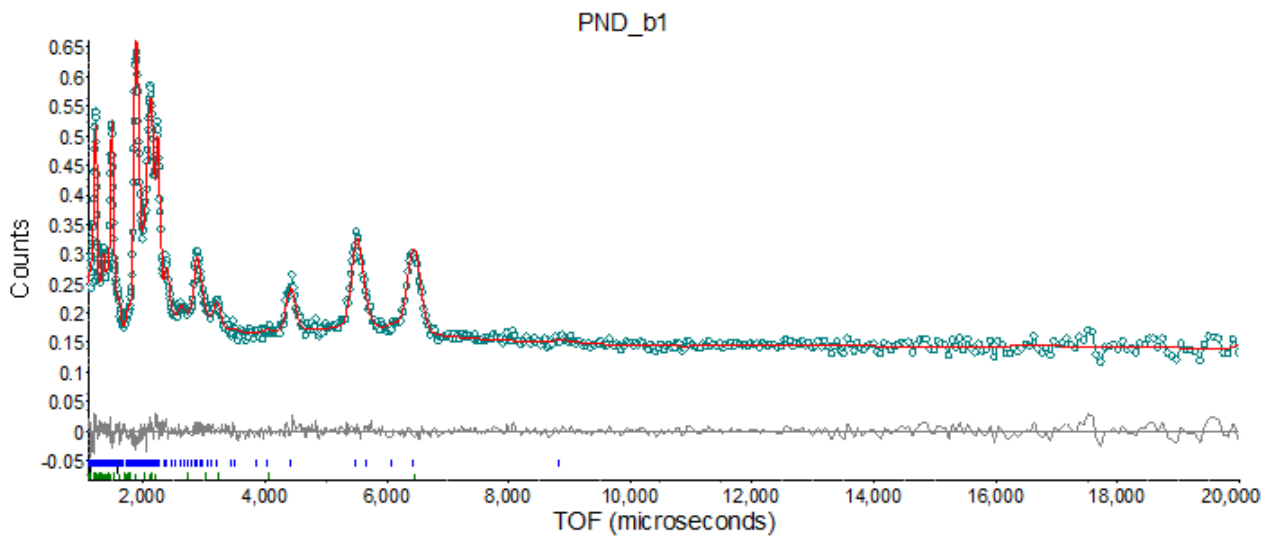
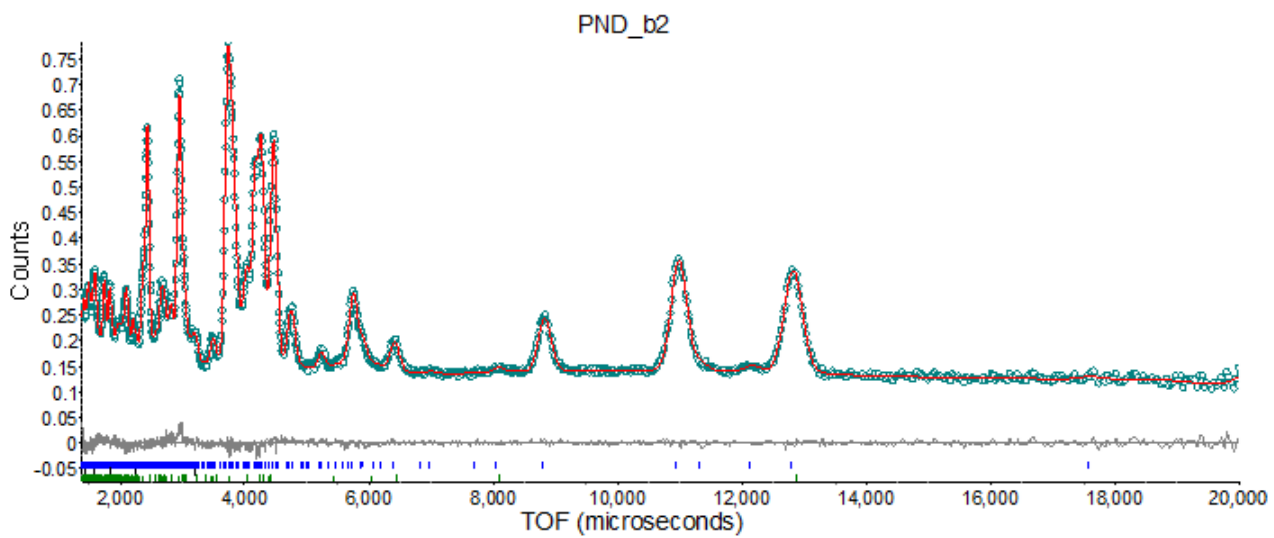
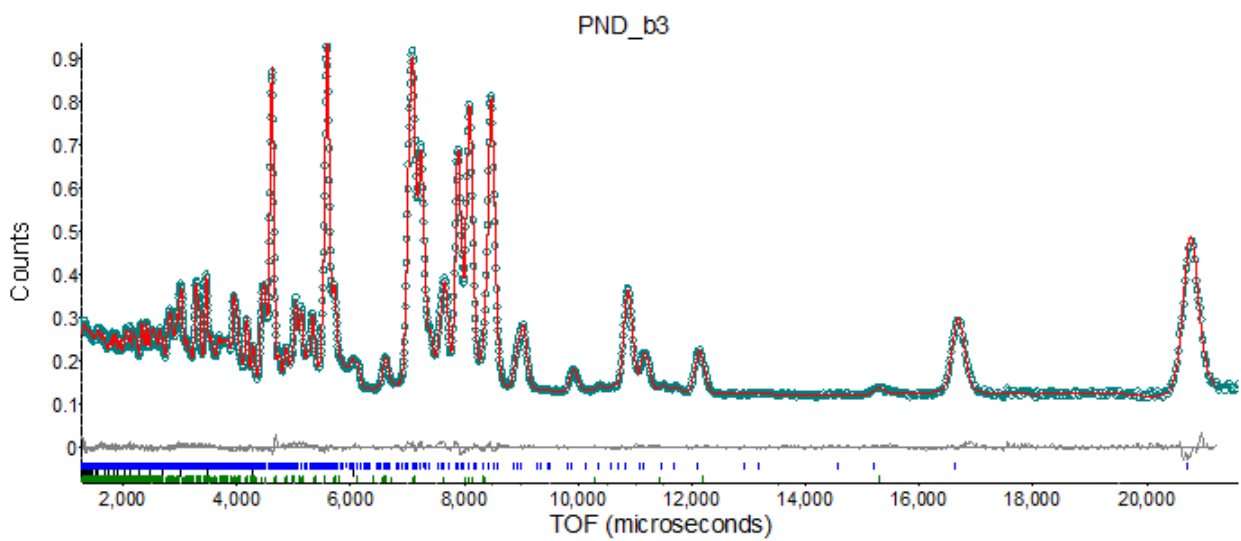

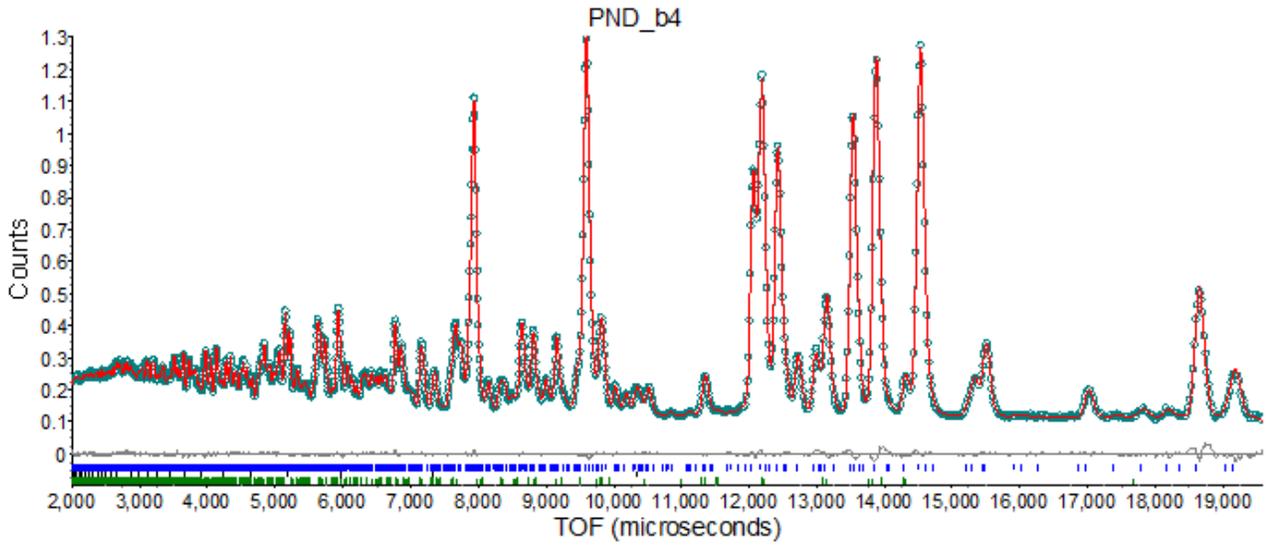
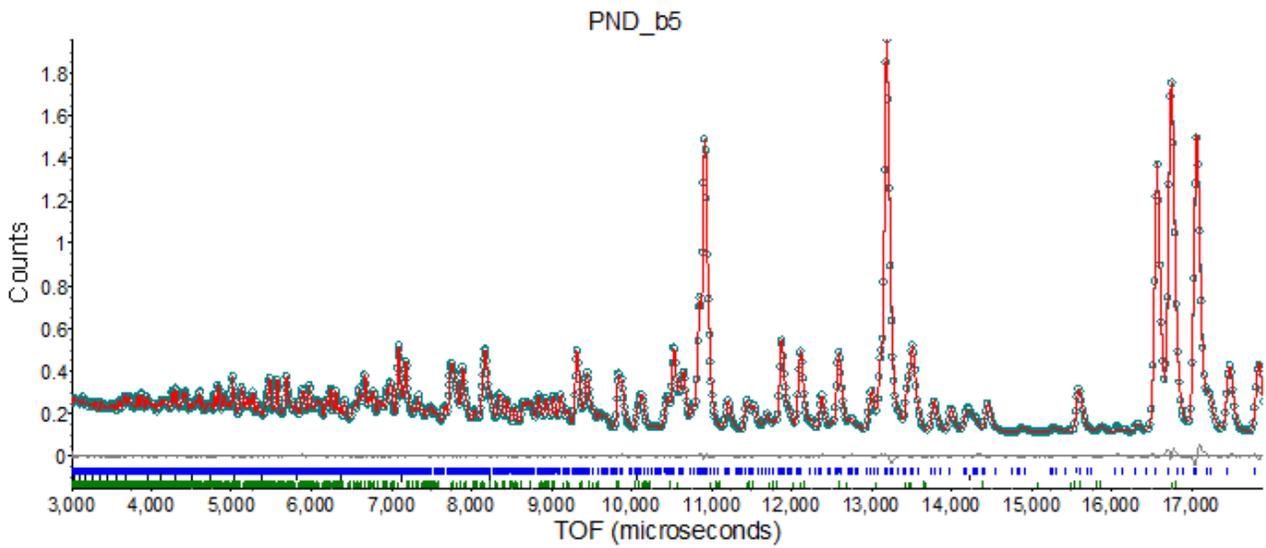
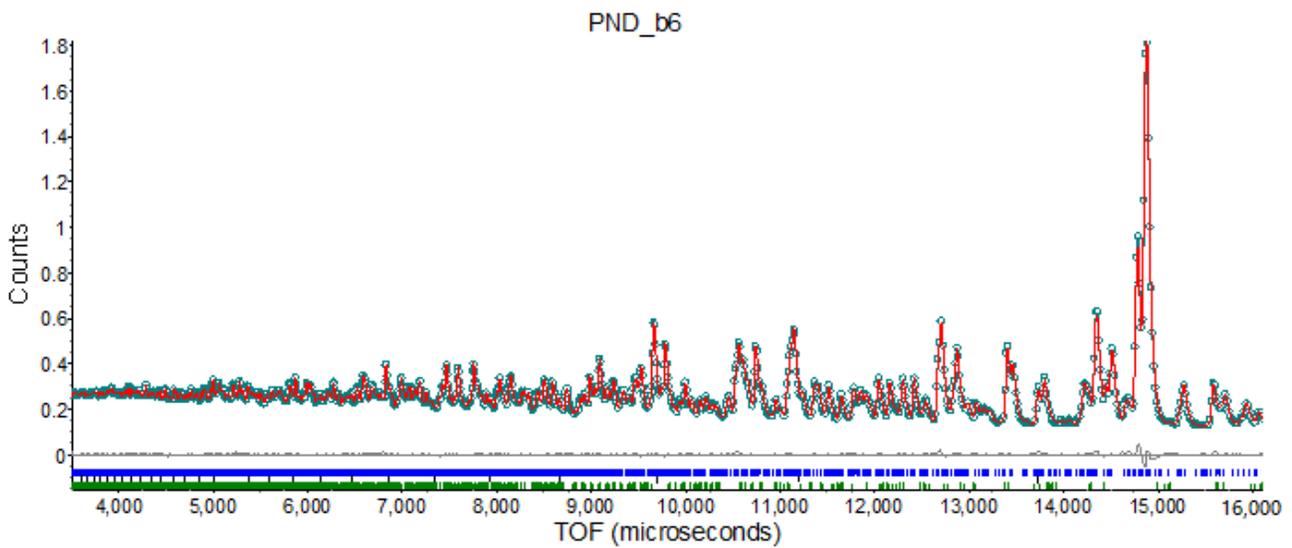

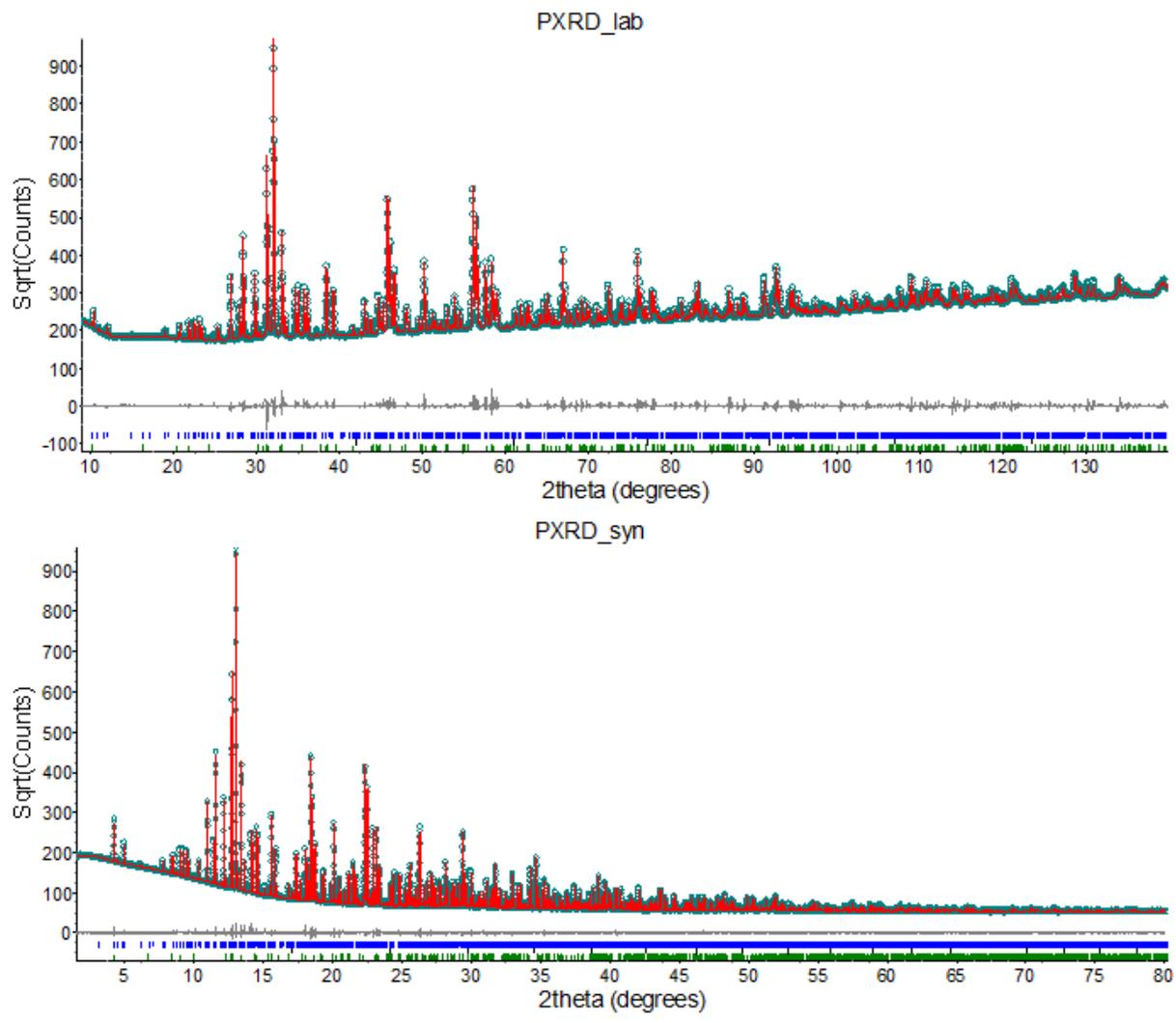

Fig. S3. Rietveld refinement profiles of $Ce_2O_2FeSe_2$ from combined refinement of room temperature data using $Pna2_1$ model. *Dots: observed, solid line: calculated curve; grey line below: difference curve; vertical tick marks: peak positions.*

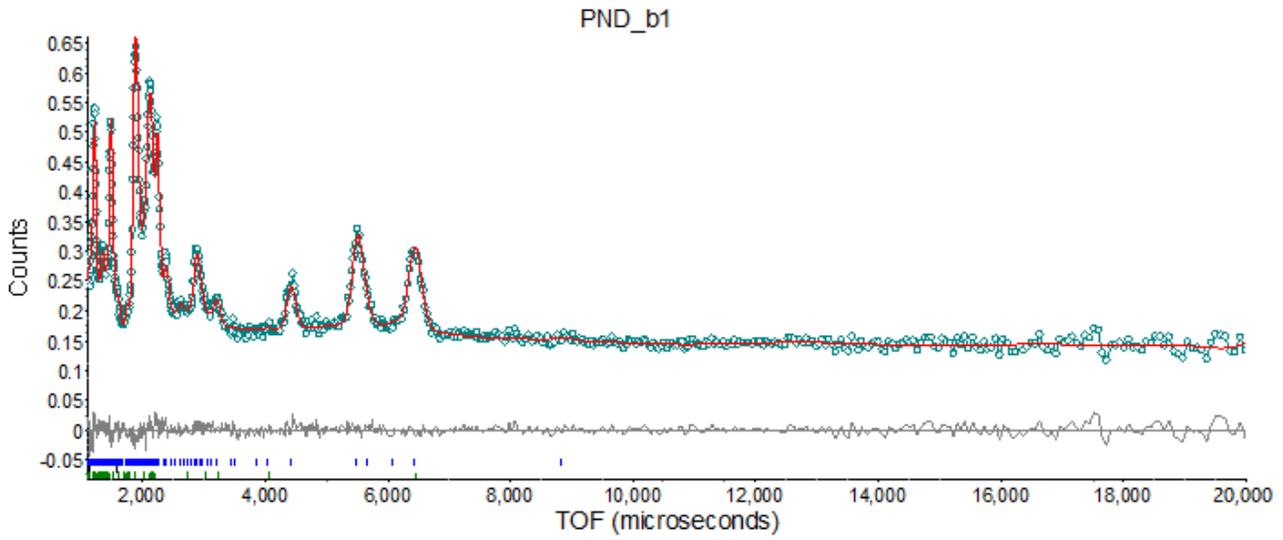
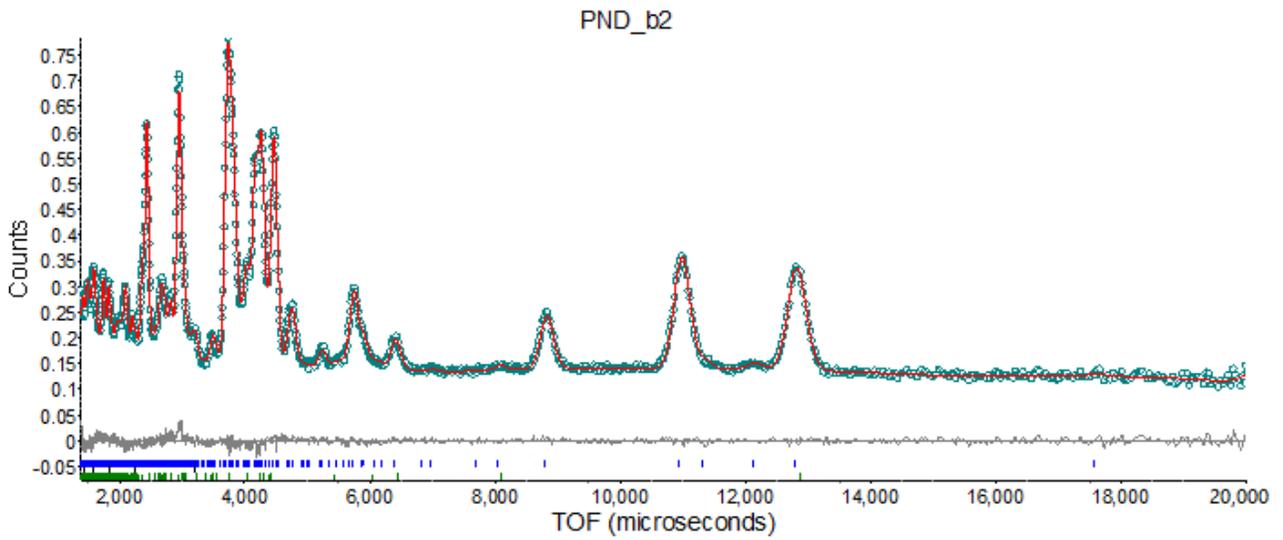
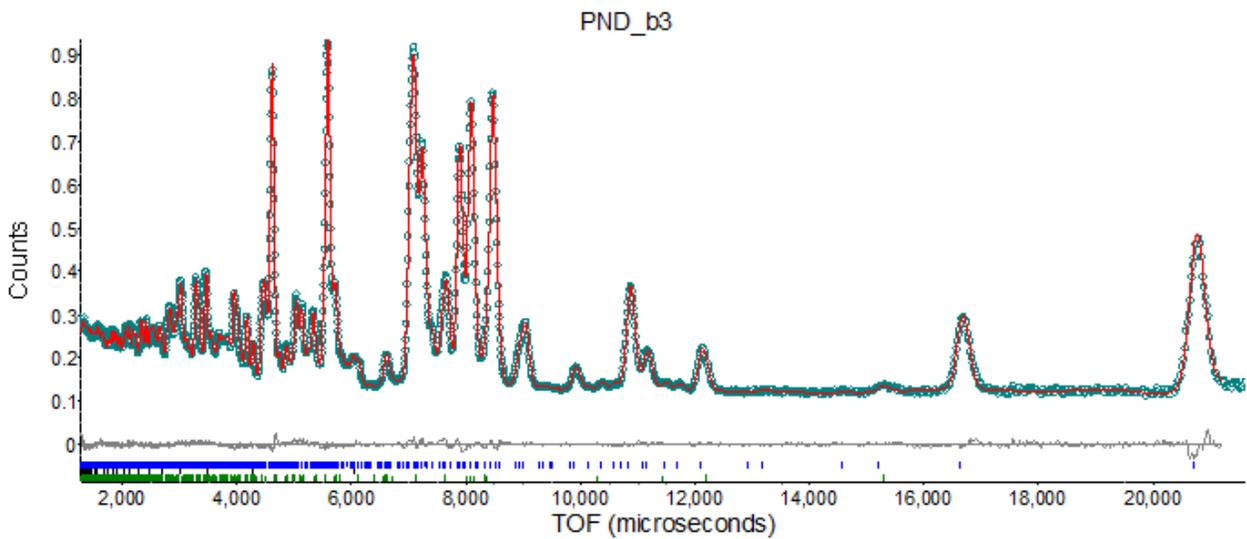

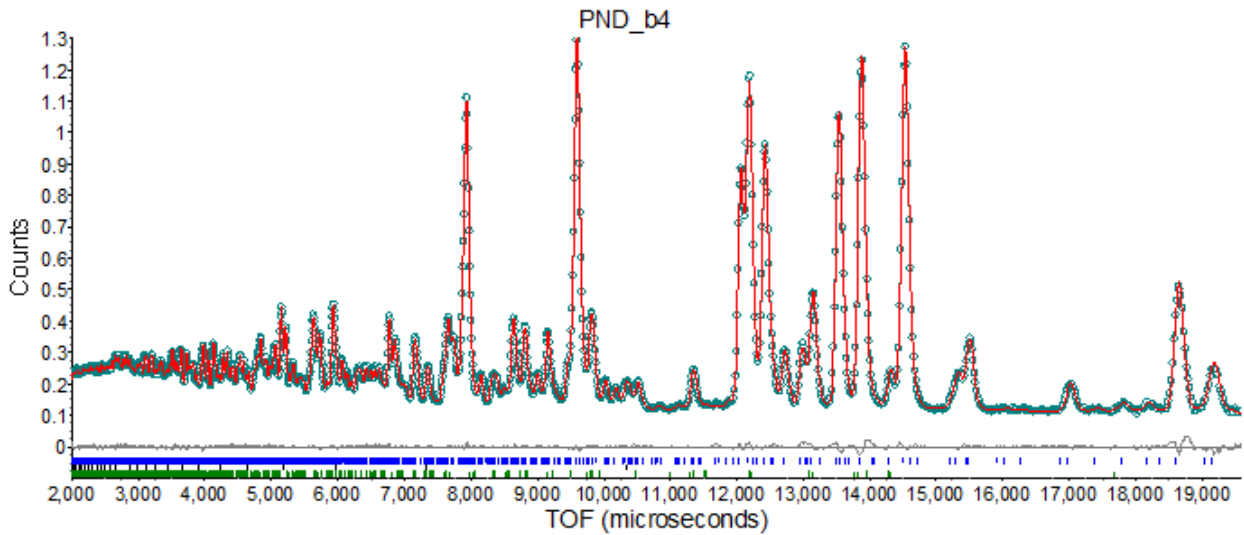
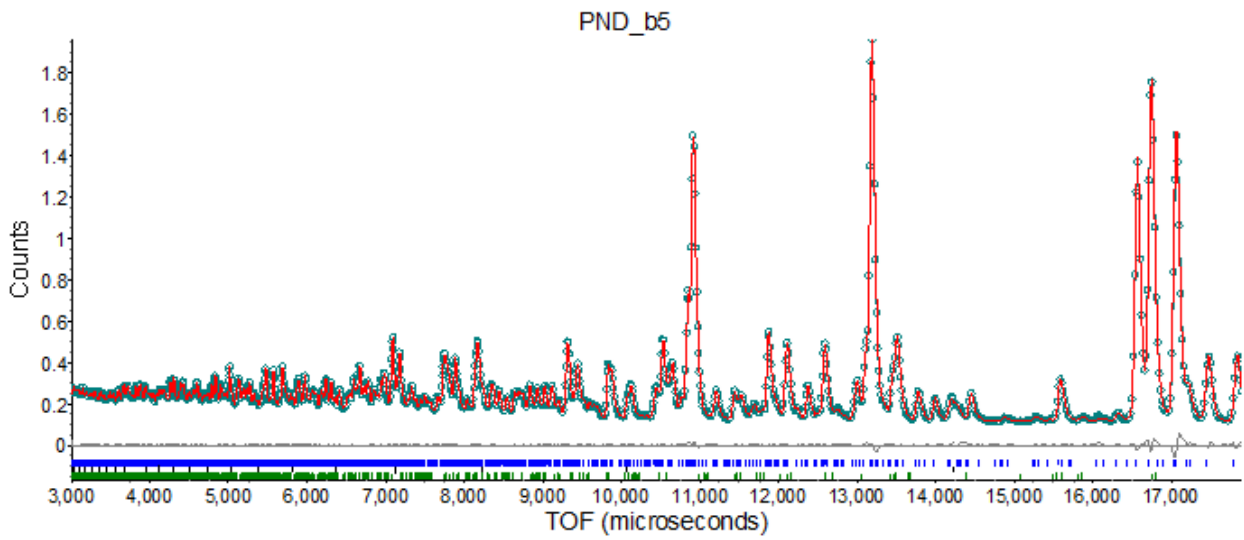
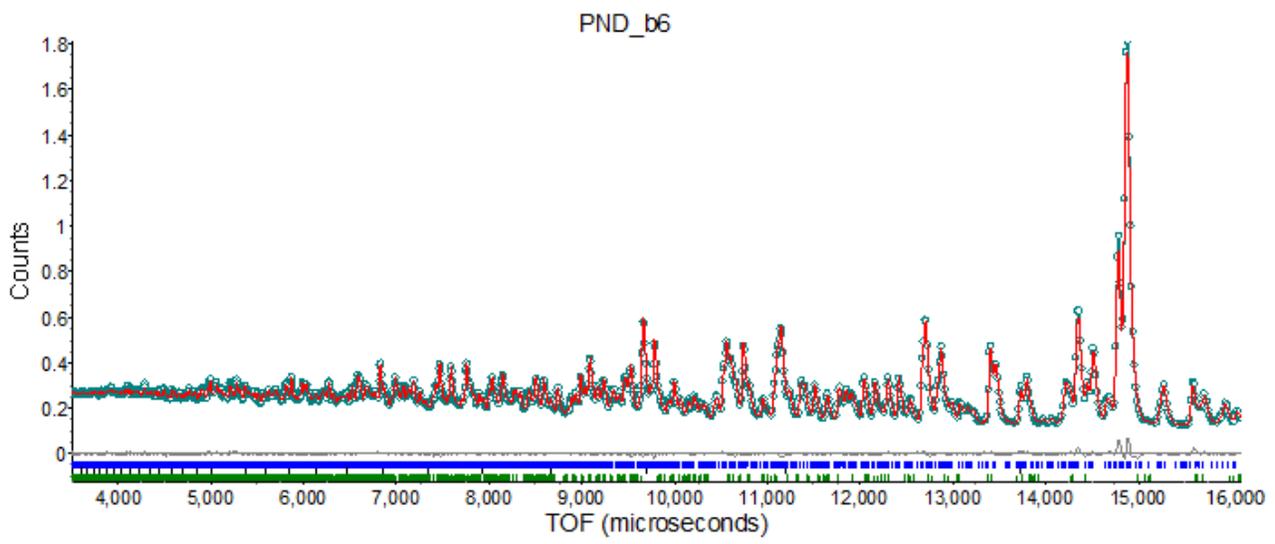

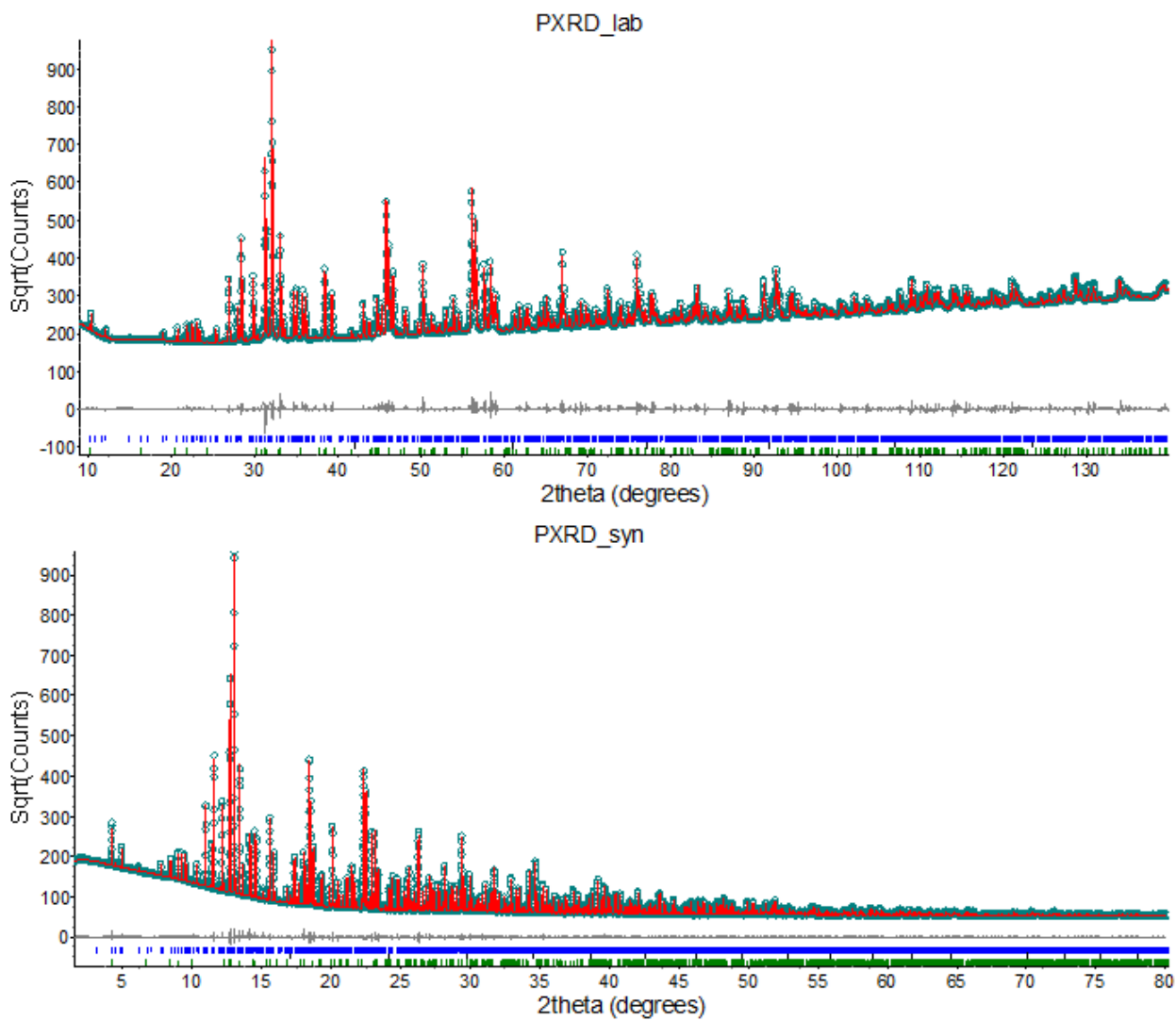

Fig. S4. Rietveld refinement profiles of Ce$_2$O$_2$FeSe$_2$ from combined refinement of room temperature data using P*nma* model.
*Dots: observed, solid line: calculated curve; grey line below: difference curve; vertical tick marks: peak positions.*

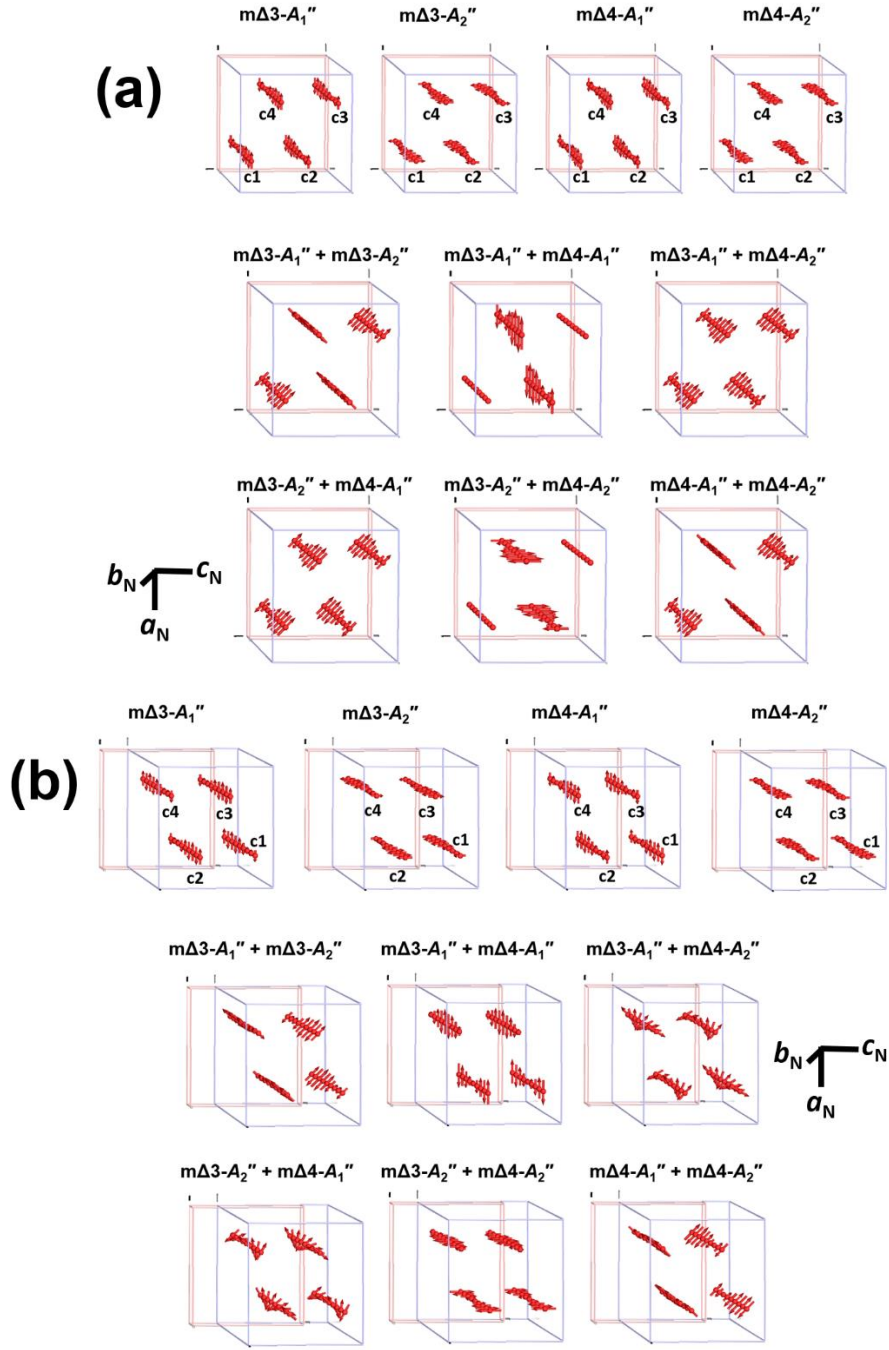

Fig. S5. The mΔ3 and mΔ4 magnetic moments modes in $a_Nc_N$ plane of Fe1 (cf nuclear structure) and their combinations. (a) and (b) are two different phase choices between mΔ3 and mΔ4; c1 to c4 indicate the different chains along $b_N$ axis and the amplitude of all modes are set to the same in the figure. The PND data was fitted best by mΔ3-$A_2''$ + mΔ4-$A_1''$ models (both choices).

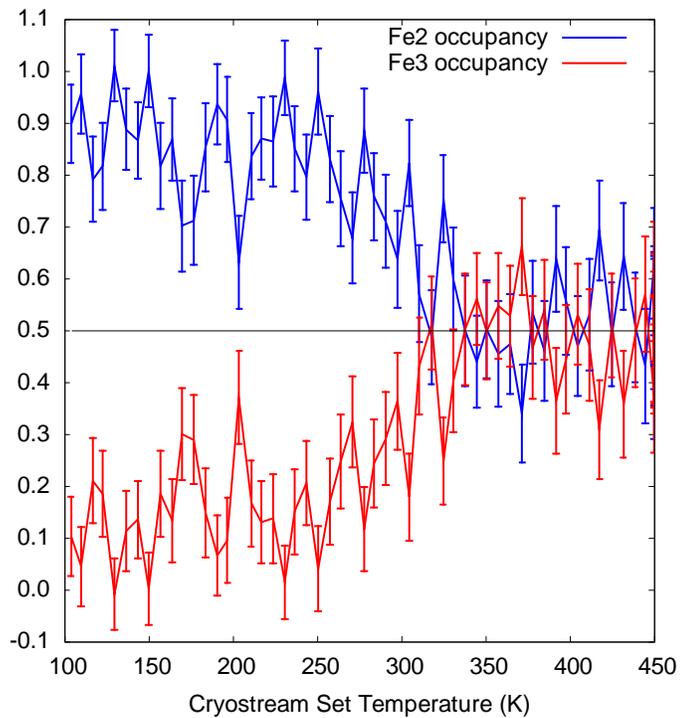

Fig. S6. Refined Fe site occupancies from "High T" variable temperature powder diffraction data discussed in the text. Data were collected relatively rapidly and the contribution of Fe2/Fe3 site ordering to diffraction data is relatively low. Despite this, the evolution of Fe site ordering with temperature can be observed.